\documentclass[12pt]{article}
\usepackage{amssymb}
\usepackage{amsmath}
\newcommand{\R}{{\mathbb R}}
\newcommand{\C}{{\mathbb C}}
\newcommand{\M}{{\mathbb M}}

\newcommand{\Z}{{\mathbb Z}}
\newcommand{\G}{\Gamma}
\newcommand{\Cl}{{{\mathbb C}\ell}}
\newcommand{\g}{\gamma}
\renewcommand{\d}{\delta}
\newcommand{\f}{\phi}
\newcommand{\vf}{\varphi}
\newcommand{\s}{\sigma}

\renewcommand{\u}{{\upsilon}}

\renewcommand{\gg}{{\mathcal G}}
\newcommand{\remove}[1]{{}}
\numberwithin{equation}{section}
\newcommand{\secs}[1]{ %\setcounter{equation}{0}
                       \section{#1}}
\begin{document}

\def\one{1\kern-3pt \text{l}}

\title{On Fermions in Compact Momentum Spaces\\
Bilinearly Constructed with Pure Spinors}
\author{Paolo Budinich\footnote{International School for Advanced Studies,
Trieste 34014, Italy}}
\date{}
\maketitle

\abstract{It is shown how the old Cartan's
conjecture on the fundamental role of the geometry of simple (or pure)
spinors, as bilinearly underlying euclidean geometry, may be
extended also to quantum mechanics of fermions (in first
quantization), however in compact momentum spaces, bilinearly constructed
with spinors, with signatures unambiguously resulting from the
construction, and locally conceived as both Fourier- and conformally-dual to
space-time, where classical mechanics is traditionally described.
In this construction most of the elementary equations of motion of
fermion-physics, usually postulated ad hoc, are naturally
obtained from Cartan's equation defining spinors. We start from
two-component spinors obeying Weyl equation for massless
neutrinos (from which in turn Maxwell equations are derived)
following with eight-component spinor-equations representing
charged-neutral fermion doublets presenting $SU(2) \otimes U(1)$
internal symmetry, including the skeleton of the electroweak
model, up to sixteen component Majorana-Weyl spinors associated with the
real Clifford algebra $\Cl(1,9)$, where, because of the known periodicity
theorem, the construction naturally
ends. $\Cl(1,9)$ may be
formulated in terms of the octonion division algebra, at the
origin of $SU(3)$ internal symmetry, and which seems appropriate
to furnish a natural explanation for several of the observed
properties of baryon- and lepton-physics.

In this approach the extra dimensions beyond 4 appear as
interaction terms in the equations of motion of the fermion
multiplet; more precisely the directions from 5$^{th}$ to
8$^{th}$  correspond to electric, weak and isospin interactions
$(SU(2) \otimes U(1))$, while those from 8$^{th}$ to 10$^{th}$ to
strong ones $(SU(3))$. Furthermore, dimensional reduction in
momentum space, which is compact, may be simply identified with ``decoupling''.
In fact it is generated by operators which
are extensions of the familiar chiral projectors in spinor space,
and they then naturally reduce both the spinor space of fermion multiplets
(reduced by a factor 1/2) and the corresponding interaction terms, of which
two are eliminated, corresponding to two extra dimensions in momentum
space, for which then there seems to be no need of the
corresponding configuration-space. Only four dimensional
space-time is needed -- for the equations of motion and for the local
fields -- and also naturally generated by four-momenta as Poincar\'e translations.

This spinor approach could be compatible with string theories and
even explain their origin, since also strings  may be bilinearly
obtained from simple (or pure) spinors through sums; that is integrals of null
vectors -- which are generalizations of the old Enneper-Weierstrass
parametrization of minimal surfaces -- in the case of real Clifford
algebras, like $\Cl(1,9)$, admitting Majorana-Weyl spinors. }

\secs{INTRODUCTION}

    In the study of the elementary constituents of matter, the
discovery of the existence of fermion- and boson-multiplets, which
could be labeled according to the representations of certain
groups ($SU(2)$, $SU(3)$...), named internal symmetry groups, has
brought to the conjecture of the existence of higher dimensional
space-times, in which to imbed ordinary space-time. The
non-observable dimensions ($> 4$) were then eliminated with
appropriate methods (e.g. Kaluza Klein), confining them in compact
manifolds of invisible size.

    Here we propose a more economical ``constructive'' approach which
simply consists in representing the observed fermions, by Pauli,
Weyl or Dirac spinors to be embedded in higher
dimensional spinor spaces and consequently the same for their
endomorphism- Clifford algebras, and, at each step of this
construction, which unambiguously defines the signatures of the
higher dimensional Clifford algebras, observe which are the
equations of motion whose geometry is contained in the Cartan's
equation defining spinors, if interpreted in momentum space.

    The main motivation of this approach is not only economy; that
is to introduce, in the attempts of theoretical explanations, only
actually observed geometrical objects that is spinors,
representing fermions, (the bosons may be notoriously expressed,
bilinearly, in terms of spinors), but rather to literally adhere
to the hypothesis of E. Cartan~\cite{1} who conjectured that the
fundamental geometry appropriate for the description and
understanding of elementary natural phenomena is spinor geometry,
more precisely the geometry of simple spinors, later named pure by C.
Chevalley~\cite{2},
rather than the one of euclidean vectors which may be constructed
bilinearly from spinors.

    After a short review on the properties of simple or pure
spinors in Chapter \ref{sec1}, we will start by considering the
elementary case of two component Pauli spinors associated with
3-dimensional euclidean space and show how we may obtain from
their Clifford algebra both the signature of space-time and Weyl
and Maxwell's equations (Chapter \ref{sec2}), in momentum space.

    We will then study in Chapter \ref{sec3} the problem of
imbedding spinor spaces and null vector spaces in higher
dimensional ones, and we will show how it may be easily solved,
in the case of simple or pure spinors, with the use of two
propositions. In Chapter \ref{sec4} we will deal with four
component spinors and with the Dirac equation for massive
fermions and the Cartan's equation for Weyl spinors or twistors.

    We will show then that the minimal Clifford algebra to contain
simple Weyl spinors isomorphic to
doublets of Dirac spinors is $\Cl(7,1)$ or  $\Cl(1,7)$ associated
with eight dimensional vector spaces and how from this both the equation
for the nucleon doublet interacting with the pseudoscalar pion
isovector may be naturally obtained, in which the pion is
bilinearly expressed in terms of the spinors of the nucleon
doublet (section \ref{sec5.1}), and also the geometrical skeleton
of the Salam-Weinberg model of electroweak interactions (section
\ref{sec5.2}); all this in momentum space.

We will then show, in section 6.3, how the equations for fermion
doublets naturally present an $U(1)$ symmetry, for one of the
fermions of the doublet, interpretable as charge, which may be
correlated with non equivalent spinor structures in conformal-like
theories and furnish then a geometrical explanation of the
existence of charged -- neutral fermion doublets like
electron-neutrino, proton-neutron, etc.

These equations, which historically have been proposed ad hoc for
the description of the mentioned phenomena, here seem to have a
unique geometrical origin in spinor geometry. They are however to
be interpreted in momentum space, thus supporting the conjecture,
formulated some time ago, that it is momentum space the
appropriate space for the geometrical formulation of quantum
mechanics (in first
quantization)~\cite{3}. And if we follow the
suggestion of Cartan to privilege simple or pure spinors, that is
maximal totally null planes laying in Klein quadrics, these
momentum spaces are compact, thus a priori eliminating the severe
difficulty of ultraviolet divergences.

The internal symmetry ($SU(2)$ in the mentioned case) that thus
arises appears to be generated, in flat spaces, by discrete
groups of reflection operators, of the conformal group. Through reflections the
heuristic approach above can be derived from the hypothesis of
the fundamental role of conformal covariance which could impose
the compact structure of phase space where space-time and
momentum space appear as conjugate with respect to conformal
reflections (Chapter {\ref{sec6}}).
%The possible genesis of gauge groups from reflections
%is dealt with in section \ref{sec6b}.

16-component spinors will be studied in Chapter 8 where a new
$U(1)$-charge appears for one of the fermion doublets, which, if
interpreted as strong charge, suggests the interpretation of the
multiplet as baryon-lepton quadruplet.

In Chapter 9 we will study 32-component spinors associated with
the algebra $\Cl(1,9)$ where the ''construction'' finds a natural
end, since after that, due to the periodicity theorem, the sequence
will be repeated.

In Chapter 10 we will discuss the problem of dimensional
reduction from $\Cl(1,9)$  which will simply consist in reversing
the steps of the ''construction'', with the use of projector
operators in spinor space. There are $3+1$ of them corresponding
to the dimensions of quaternion-space which, as shown in Chapter
11, section 11.3, might explain the geometrical origin of the
''families''. In Chapter 12 the octonion formalism is introduced
in the frame of $\Cl(1,9)$  to show that it may give rise to
several $SU(3)$ internal symmetry subgroups of $G_2$ in the equations of the
baryon
multiplet. Further geometrical aspects are mentioned in Chapter
13, among these simplicity constraints in section 13.1.

The present approach, rather than alternative to the now
prevailing approach based on strings, could be compatible with it,
since strings may be be formulated as integrals of null-vectors
which are notoriously the characteristic transition elements (bilinear) from
spinor to euclidean geometry. Spinors then could be at the
origin of strings insofar these could merely represent the intermediate stage between spinor- and euclidean-geometry as mentioned in section~13.2.

\secs{SIMPLE- OR PURE-SPINOR GEOMETRY}\label{sec1}

The geometry of simple or pure spinors was discovered by
\'E.~Cartan. He formulated the basic axioms and theorems which, for
what concerns us, may be summarized~\cite{4} as follows.

    Given a $2n$-dimensional complex space $W= \C^{2n}$ and the
corresponding central simple Clifford algebra $\Cl(2n)$ with generators
$\g_a$, $(a = 1,\dots,2n)$, obeying $\left\{\g_a, \g_b\right\} =
(-1)^{a+1} 2\d_{ab}$, a spinor $\phi$ is a vector of the complex $2^n$
dimensional representation space $S$ of $\Cl(2n) = {\rm End} S$,
defined by
\begin{equation}
    z_a \g^a \phi = 0   \label{1.1}
\end{equation}
where $z_a$ are the orthonormal components of a vector $z \in W$. This
vector is null since, for $\phi \not= 0$, eq. (\ref{1.1}) defines the Klein
quadric $Q \subset W$:
\begin{equation}
    Q \colon \qquad z_a z^a = 0 \label{1.2}
\end{equation}

For fixed $\f$, all $z \in W$ satisfying (\ref{1.1}) and (\ref{1.2})
define a totally null plane $T(\f) \subset W$, whose vectors are both
null and mutually orthogonal.

Let $\g_{2n+1}= \g_1 \cdot \g_2 \cdots \g_{2n}$ represent the volume
element (normalized to 1) of $\Cl(2n)$, then the spinors $\f_+$,
$\f_-$ obeying
\begin{equation}
    \g_{2n+1} \f_\pm = \pm \f_\pm   \label{1.3}
\end{equation}
are named Weyl spinors and for them the defining equation (\ref{1.1})
becomes
\begin{equation}
    z_a \g^a (1 \pm \g_{2n+1}) \f_\pm = 0   \label{1.4}
\end{equation}

A Weyl spinor $\f_+$ or $\f_-$ is named simple or pure if the associated
totally null plane $T(\f_\pm)$ is n-dimensional, that
is, maximal and we will write $T(\phi_\pm )=M(\phi_\pm )$.  For $n \leqslant 3$
all Weyl spinors
$\f_+$,
$\f_-$ are simple.

The $2^{n-1}$ dimensional spaces $S_+$ and $S_-$ of Weyl spinors are
endomorphism spaces of the even Clifford subalgebra $\Cl_0(2n)$ of
$\Cl(2n)$ that is:
\begin{equation}
  \Cl_0(2n) = {\rm End} S_\pm . \label{1.5}
\end{equation}
We have further:
\begin{equation}
  \f = \f_+ \oplus \f_- \; , \qquad S = S_+ \oplus
S_- . \label{1.6}
\end{equation}
and $\Cl(2n) = 2 \Cl_0(2n)$.

    These definitions may easily be extended also to odd dimensional
spaces, in fact, since the volume element
$\gamma_{2n+1}$, for every $\gamma_a$, obey to:
$\left\{\g_a, \g_{2n+1}\right\} = 2\d_{a,
2n+1}$, $1\leq a\leq 2n$ we have that $\g_1$, $\g_2$, ...
$\g_{2n}$, $\g_{2n+1}$ generate the Clifford algebra $\Cl(2n+1)$
of the complex vector space $W = \C^{2n+1}$ and there is the
isomorphism~\cite{4}:
\begin{equation}
    \Cl_0(2n+1) \simeq \Cl(2n)  \label{1.7} %\tag{\ref{1.7}$'$}
\end{equation}
both being simple. The corresponding $2^n$ component spinors are
called Pauli spinors: $\f_P$, for $\Cl_0(2n+1)$ and Dirac spinors:
$\f_D$ for $\Cl(2n)$.

$\Cl(2n+1)$ instead is not simple and there is the isomorphism [4]:
\begin{equation}
    \Cl(2n+1) \simeq \Cl_0(2n+2)  \label{1.7p} \tag{\ref{1.7}$'$}
\end{equation}

For embedding spinors in higher dimensional spinors we may then use
eqs. (\ref{1.5}), (\ref{1.6}), (\ref{1.7}) and (2.7$'$):
\begin{eqnarray*}
 \Cl(2n) \simeq \Cl_0(2n +1) \hookrightarrow  \Cl(2n+1)
 \simeq \mspace{60mu} \\ \simeq \Cl_0(2n+2) \hookrightarrow  \Cl(2n+2) \dots
\end{eqnarray*}
corresponding to the spinor embeddings:
\begin{equation*}
  \psi_D \simeq \psi_P \hookrightarrow \psi_P \oplus\psi_P \simeq
  \psi_W^+ \oplus \psi_W^- = \Psi =\psi_D \oplus \psi_D
\end{equation*}
which means that a $2^n$ components Dirac spinors may be
considered equivalent to a doublet of $2^{n-1}$ components
Weyl,Pauli or Dirac spinors.  For simple spinors, these
embeddings will correspond to embeddings of Klein quadric in
higher dimensional ones which, when restricted to the real and
projective quadrics, will correspond to the embedding of compact
manifolds in higher dimensional compact manifolds.

The vectors (and tensors) of the manifolds, to be interpreted as
vector spaces of physics, will result bilinearly expressed in terms of
spinors, therefore we need to define inner products in spinor
spaces. This is easily done through the main anti-automorphism of the
simple Clifford algebras $\Cl(2n)$ or $\Cl_0(2n+1)$ which defines an
isomorphism $B : S \rightarrow S^\ast$ where $S^\ast$ is the dual
spinor space of $S$ such that
\begin{equation}
B \g_a = \g_a^{t} B \quad {\rm and} \quad B \f = \f^t B \in S^\ast
\label{1.8} \end{equation} where $\g_a^t$ and $\f^t$ mean $\g_a$
and $\f$ transposed, respectively.

If $\psi$ is another spinor of $S$ we have then the invariant (for the
$Pin$ group) scalar product:
\begin{equation*}
 \langle \f^\ast , \psi \rangle = \langle B \f , \psi \rangle = \f^t B
 \psi.
\end{equation*}

In the case of real vector spaces, of interest for physics, we will also need to define the conjugation operator $C$
such that
\begin{equation}
  C \g_a = {\bar \g_a} C \quad {\rm and } \quad \f^c = C \bar \f
  \label{1.9}
\end{equation}
where $\bar \g_a$ and $\bar \f$ mean $\g_a$ and $\f$ complex
conjugate, respectively.

Another useful definition of simple spinors may be obtained through
the formula
\begin{equation}
\f \otimes B \psi = \sum\limits_{j=0}^n F_j \label{1.10}
\end{equation}
where $\f, \psi \in S$ are spinors of $\Cl(2n) = {\rm End} S$ and
\begin{equation}
  F_j = {}_[\g_{a_1} \g_{a_2} \cdots \g_{a_j}{}_] T^{a_1 a_2 \dots
  a_j} \tag{\ref{1.10}$'$} \label{1.10p}
\end{equation}
where the $\g_a$ products are antisymmetrized and $ T^{a_1 a_2 \dots a_j}$
is an antisymmetric $j$-tensor of $\C^{2n}$, which can be expressed
bilinearly in terms of the spinors $\f$ and $\psi$ as follows:
\begin{equation}
  T_{a_1 a_2 \dots a_j}= \frac{1}{2^n} \langle B \psi, {}_[ \g_{a_1}
  \g_{a_2} \cdots \g_{a_j }{}_] \f \rangle \label{1.11}
\end{equation}
Setting now $\psi = \f$, in eq.~(\ref{1.10}) we have that $\f$ is
simple if and only if
\begin{equation}
  F_0=0,\quad F_1 = 0, \quad F_2 = 0,\dots , F_{n-1} = 0
  \label{1.12}
\end{equation}
while $F_n \not= 0$ and eq.~(\ref{1.10}) becomes:
\begin{equation}
   \phi \otimes B \phi = F_n \label{1.13}
\end{equation}
and the $n$-tensor $F_n$ represents the maximal totally null
plane of $W$ equivalent, up to a sign, to the simple spinor $\f$.
Equations~(\ref{1.12}) represent then the constraint equations
for a spinor $\f$ to be a simple or pure spinor associated with $W$.

Eq.~(\ref{1.13}) represents then the correspondence of simple or pure
spinor directions with maximal totally null planes sometimes
called ``the Cartan map''.

The equivalence of this definition with the one deriving from
eq. (\ref{1.1}), given by Cartan, is easily obtained if we multiply
eq.~(\ref{1.10}) on the left by $\g_a$ and on the right by $\g_a\f$
and sum over $a$, obtaining
\begin{equation}
    \g_a \f \otimes B \psi \g^a \f = z_a \g^a \f    \label{1.14}
\end{equation}
where
\begin{equation*}
z_a = \langle B \psi, \g_a \f \rangle \label{1.15}
\end{equation*}
which, provided $\f$ is simple or pure , for arbitrary $\psi$, satisfy
\begin{equation*}
  z_a \g^a \f = 0.
\end{equation*}
and $z_a$ are the components of a null vector of $W$, belonging to
$F_n$.

This result may be formulated as follows:\\

\noindent \underline{Proposition 1.}\ Given a complex space $W=\C^{2n}$ with
its Clifford algebra $\Cl (2n)$, with generators $\gamma_a$, let $\psi$ and
$\phi$ represent two spinors of the endomorphism spinor-space of $\Cl (2n)$
and of its even subalgebra $\Cl_0(2n)$, respectively. Then, the vector $z\in W$,
with components:
\begin{equation}
z_a =\langle B\psi ,\gamma_a\phi\rangle ;\quad a=1,2,\dots 2n
\end{equation}
is null if and only if $\phi$ is a simple or pure spinor. For fixed $\phi$
simple or pure, and arbitrary $\psi$ all vectors of the maximal,
totally null plane $M(\phi )$ in $W$, are so obtained.
The proof is given in reference [6].

    The above formalism may be easily restricted to the real, simply
substituting the complex space $W = \C^{2n}$ with the real neutral
pseudo-euclidean space $V = \R^{n,n}$. The corresponding Clifford algebra
$\Cl(n,n)$ is generated by the generators $\g_a$, satisfying:
\begin{equation}
    \left\{\g_a, \g_b\right\} = (-1)^{a+1}\ 2\d_{ab}\label{1.16}
\end{equation}
and, in the previous computations the complex components $z_a$ of the
vector $z \in W$ have to be substituted by real ones $p_a$ of $p \in
V$, and in eqs.~(\ref{1.10}), (\ref{1.11}), (\ref{1.14}) and
(\ref{1.15}), $B \f$ and $B \psi$ have to be substituted by $B \f^c$
and $B \psi^c$, respectively.

    The same formalism may be also extended~\cite{5} to real
pseudo-euclidean spaces $\M = \R^{n+1,n-1}$ which, for $n = 2$,
identifies with Minkowski space-time and, for $n > 2$, represents
its conformal extensions, as well as of its Fourier dual momentum
space. In this case $z_a$ are real (or imaginary) only for $n$
even, that is for $\Cl(3,1)$, $\Cl(5,3)$, $\Cl(7,5)$ etc. (while
they are complex for $n$ odd, that is for $\Cl(4,2)$, $\Cl(6,4)$
etc.)

In fact in this case, of interest for physics we have:\\

\noindent \underline{Corollary 1.}\ Let $\Cl(n+1, n-1) = {\rm End}
S$, the vectors with components
\begin{equation}
   z_a^\pm = \langle B \psi^c, \g_a \f^\pm \rangle \label{1.17}
\end{equation}
where $\psi$ is an arbitrary spinor of $S$, are null iff $\f^\pm$ are simple
Weyl spinors of $S$. For $\psi = \f^\pm$ and $n$ even $z^\pm_a = p^\pm_a$
are real (or imaginary), for $n$ odd ${z_a}^\pm$ are complex such that
${{\bar z}_a}^+ = \pm {z_a}^-$.

It is seen that eq.~(2.17) is a particular case of eq.~(2.15).
In reference~\cite{6}, it is also shown
that real vectors are obtained in the case of Lorentzian
signature: $\Cl(2n-1,1)$.

For $n$ even then, $p_a$ given by (\ref{1.17}) define the Klein quadric
$Q$ given by eq.~(\ref{1.2}) where $z_a$ is substituted by $p_a$, in
$V = \R^{n+1,n-1}$.

It is easily seen that the corresponding projective quadric $PQ$
define the following compact manifold $PQ$ in $\R^{n+1,n-1}$:
\begin{equation}
  PQ = \frac{S^n \times S^{n-2}}{\Z_2} . \label{1.18}
\end{equation}
where $\Z_2 =[+1,-1]$ means that the antipodal points of $P Q$
are to be identified.

We have now listed the main geometrical instruments of simple or
pure  spinor geometry useful in order to proceed with the program
of imbedding spinor spaces in higher dimensional spinor spaces and
to explore which geometrical objects of possible physical meaning
we might obtain through this procedure.

    We will start with the simplest non trivial case of two component
spinors and then proceed to higher ones and we will see that we
will obtain, together with Maxwell's equations also most of the
elementary equations of fermion physics in momentum space known to
us, and nothing else. In other words it appears that every
one of the geometrical structures potentially contained in simple or pure
spinor geometry are realized in some elementary law of physics
supporting thus the Cartan's conjecture of the fundamental role
of simple spinor geometry not only for euclidean geometry but
also for physics of fermions the most elementary constituents of
matter, whose equations of motion already contain some of the
geometrical elements of quantum physics (in first quantization:
the non-relativistic limit of Dirac equation is Schr\"odinger
equation, up to the definition of the Plank's constant).

\secs{FROM TWO TO FOUR COMPONENT SPINORS.}\label{sec2}

\subsection{\ THE SIGNATURE OF SPACE- TIME \ AND WEYL EQUATIONS FOR MASSLESS
NEUTRINOS}

Let us start from $W = \C^3$, the generators $\s_1$, $\s_2$, $\s_3$ of
its Clifford algebra $\Cl(3) = {\rm End} S$, are Pauli matrices and
its Pauli spinors $\vf =
\begin{pmatrix} \vf_0 \cr \vf_1 \end{pmatrix}
\in S$ are simple.\footnote{Because of the isomorphism $\Cl_0(3)\simeq
 \Cl(2)$ the two components spinors $\varphi$ may be also conceived as
a Dirac spinor associated with $V=\C^2$, which however may not be
simple: in fact $\langle B \vf, \s_3 \vf \rangle \not= 0$. In
general only Pauli or Weyl spinors may be simple, Dirac ones may
not (unless they are conceived as isomorphic to Pauli or Weyl spinors in force
of the isomorphisms (2.7), (2.7$'$)). This is presumably the reason why in the
introduction of spinor geometry Cartan preferred odd dimensional spaces, in
particular
$\C^3$, to even dimensional ones. } In fact we have
$B= -i \sigma_2 = \begin{pmatrix} 0 & -1 \cr 1 & 0
\end{pmatrix}$ and eq.~\eqref{1.13} becomes:
\begin{equation}
  \begin{pmatrix} \vf_0 \vf_1 & - \vf_0 \vf_0 \cr \vf_1 \vf_1 & -\vf_0
  \vf_1 \end{pmatrix}\equiv \vf \otimes B \vf = z_j \s_j \equiv
  \begin{pmatrix} z_3 & z_1 -i z_2 \cr z_1+ i z_2 & -z_3 \end{pmatrix}
  \label{2.1}
\end{equation}
since $F_0=\langle B \vf, \vf \rangle \equiv 0$. Furthermore, since
$\frac{1}{2}\langle B \vf, \s_j \vf\rangle$ $= z_j$ (compare the
two matrices), following from eq.~\eqref{2.1}, equation:
\begin{equation}
     z_1^2 + z_2^2 + z_3^2 = 0   \label{2.2}
\end{equation}
is identically satisfied as may be immediately seen from the determinant of
the matrices in eq.~\eqref{2.1}. Also the Cartan's equation:
\begin{equation}
z_j \s^j \vf =0 \label{2.3}
\end{equation}
is identically satisfied, (as may be immediately seen if we act
with the first term of eq.~\eqref{2.1} on $\vf$).

If $\psi \in S$ is another spinor we have, from eq.(2.10):
\begin{equation}
  \begin{pmatrix}
  \vf_0 \psi_1 & - \vf_0 \psi_0 \cr \vf_1 \psi_1 & -\vf_1
  \psi_0 \end{pmatrix}\equiv \vf \otimes B \psi = z_0 + z_j \s_j \equiv
  \begin{pmatrix} z_0 +z_3 & z_1-iz_2 \\ z_1+iz_2 & z_0 - z_3
  \end{pmatrix} ,\label{2.4}
\end{equation}
and $z_0 = \frac{1}{2}\langle B \psi, \vf\rangle$, $z_j =
\frac{1}{2}\langle B \psi,\s_j \vf\rangle$ deriving from it,
satisfy identically the equation (as may be immediately seen from the determinant of the matrices):
\begin{equation}
   z_1^2 + z_2^2 + z_3^2 - z_0^2 = 0  \label{2.5}
\end{equation}
which uniquely determines the signature of Minkowski space-time. In fact
the above may be easily restricted to the real by substituting $B\psi$ with
$B\phi^c = \phi^\dagger$ by which $z_0$, $z_j$ become $p_0$, $p_j$
real:
\begin{equation}
    p_0 = \frac{1}{2}\langle\vf^\dagger, \vf\rangle ; \qquad p_j =
\frac{1}{2}\langle
\vf^\dagger, \s_j \vf \rangle \label{2.6}
\end{equation}
satisfying identically to:
\begin{equation}
    p_1^2 + p_2^2 + p_3^2 - p_0^2 = 0 \label{2.7}
\end{equation}
a null vector or light ray of Minkowski space $\R^{3,1}$. In this case then
exploiting the Clifford algebras isomorphism
\begin{equation}
    \Cl(3) \simeq \Cl_0(3,1) \label{2.8}
\end{equation}
$\vf$ may be interpreted as a simple Weyl spinor of $S$ where $\Cl_0(3,1) =
{\rm End} S_\pm$; and there are two of them: $\vf_+$, $\vf_-$ satisfying the
equations
\begin{equation}
\left( \vec p \cdot \vec \s + p_0 \right) \vf_+=0 \qquad \left( \vec p
\cdot \vec \s - p_0 \right) \vf_-=0 \label{2.9}
\end{equation}

These equations, and the ones in the following sections, may be
interpreted as field equations if we define the Clifford algebras as
fibers over momentum basis. For $p_0 = 0$ eqs.~\eqref{2.9} identify
with eq.~\eqref{2.3} and, in agreement with the isomorphism
\eqref{1.7} the Weyl spinors $\vf_\pm$ associated with $\Cl_0(3,1)$
identify with the Pauli spinor $\vf$ associated with $\Cl_0(3)$, apt
to represent a fermion spin in non relativistic phenomena.

Eqs.(3.9) may be expressed as a single equation for the four
component Dirac space-time spinor $\psi =\vf_+ \oplus \vf_-$. In
fact indicating with
\begin{equation*}
    \g_\mu = \left\{\g_0; \g_1, \g_2, \g_3\right\}:=
    \left\{-i\sigma_2\otimes 1; \sigma_1\otimes \sigma_j\right\} ;\
    \ j=1,2,3
\end{equation*}
the generators of $\Cl(3,1)$ and with $\g_5 = -i\g_0 \g_1 \g_2 \g_3 =
\s_3 \otimes 1$ the volume element, we may write (\ref{2.9}) in the form
\begin{equation}
  p^\pm_\mu \g^\mu \left( 1 \pm \g_5\right) \psi = 0
    \tag{\ref{2.9}$'$} \label{2.9p}
\end{equation}
where $\tfrac{1}{2} \left(1 \pm \g_5 \right)$ represent the chiral
projectors. These equations are the Weyl equations for massless
neutrinos in momentum space. The null vectors $p^\pm_\mu$ may be
expressed in the form:
\begin{equation}
p_\mu^\pm = \frac{1}{2}{\tilde \psi} \g_\mu \left(1 \pm \g_5\right) \psi
\label{2.10}
\end{equation}
where ${\tilde \psi} = \psi^\dagger \g_0$.

The space-time Weyl spinors
\begin{equation}
    \vf_\pm = \frac{1}{2}\left(1 \pm \g_5\right) \psi \label{2.11}
\end{equation}
represent massless neutrinos, eigenstates of $\g_5$.

For future
use let us now compute the components of their intrinsic angular
momentum (in units of $\frac{\hbar}{2}$) with respect to the
$z$-axis in space; which is represented by $\frac{1}{2} \s_3$,
therefore, since
\begin{equation}
  \frac{1}{2} \left( \g_5 + 1 \otimes \s_3 \right) =
  \begin{pmatrix} +1 & {} & {} & {} \cr
                  {} & 0  & {} & {} \cr
                  {} & {} & 0  & {} \cr
                  {} & {} & {} & -1
  \end{pmatrix} , \label{2.11p} \tag{\ref{2.11}$'$}
\end{equation}
it will be $\pm \frac{\hbar}{2}$ for $\vf_\pm$.

>From eq.~(3.9$'$) we may also easily obtain Maxwell's equations in
momentum space.

\subsection{MAXWELL'S EQUATIONS}

The Weyl space-time spinor $\vf_+$ is simple and therefore with it
eq. (\ref{1.13}) becomes:
\begin{equation}
  \vf_+ \otimes B \vf_+ = \frac{1}{2} F_+^{\mu\nu} \left[ \g_\mu,
  \g_\nu \right] \left( 1+\g_5\right) \label{2.12}
\end{equation}
where the antisymmetric tensor
$F^{\mu\nu}_+$ may be expressed bilinearly in terms of spinors through
eq.~(\ref{1.11}):
\begin{equation}
  F_+^{\mu\nu} = \tilde \psi \left[ \gamma^\mu, \gamma^\nu \right]
  \left( 1+ \gamma_5 \right) \psi \,,  \label{2.13}
\end{equation}
and,  as already observed by \'E. Cartan~\cite{1}, it has the geometrical
properties of the electromagnetic tensor. Now from
Weyl equation $p_\rho \g^\rho \vf_+ = 0$ we obtain from (\ref{2.12}):
\begin{equation}
p_\rho \g^\rho F^{\mu\nu}_+ \left[ \g_\mu, \g_\nu \right] \left( 1 + \g_5
\right) =0
\end{equation}
and since
\begin{equation*}
\g_\rho \g_\mu = g_{\rho\mu} + \frac{1}{2} \left[ \g_\rho, \g_\mu \right]
\end{equation*}
it becomes
\begin{equation}
  p_\rho F_+^{\rho\nu} \g_\nu = 0
\end{equation}
which is the image in $\Cl(3,1)$ of the Maxwell's equation (in vacuum) for
the self dual electromagnetic tensor in momentum space $\R^{3,1}$:
\begin{equation}
    p_\rho F_+^{\rho\nu} = 0 .   \label{2.14}
\end{equation}
It is easily seen that if we start from the left-handed Weyl
spinor $\vf_-$ we obtain the other equation:
\begin{equation}
  \varepsilon^{\lambda\rho\mu\nu} p_\rho F^-_{\mu\nu}=0
  \label{2.15}
\end{equation}
Also the Maxwell's equations in presence of electromagnetic sources may be
easily obtained from simple spinor geometry~\cite{7}.

Observe that from eq.~\eqref{2.13} it appears that the electromagnetic
tensor $F_{\mu\nu}$ is bilinearly expressed in terms of the Weyl
spinors $\vf_+$ and $\vf_-$ obeying the equation of motion
eq.~(3.9$'$) of massless neutrinos. This however does not imply that
in the quantized theory the photon must be conceived as a bound state
of neutrinos.  In fact it is known that the neutrino theory of
light, while violating both gauge invariance
and statistics, is unacceptable [8].

\secs{IMBEDDING SPINOR SPACES \\ AND NULL-VECTOR SPACES IN \\ HIGHER
DIMENSIONAL ONES} \label{sec3}

We have seen how two component spinors $\vf = \begin{pmatrix} \vf_0 \cr
\vf_1 \end{pmatrix}$ may be interpreted as Pauli spinors of $\Cl(3)$ or,
as Weyl spinors of $\Cl_0(3,1)$. For these the Cartan's equations
identify with Weyl equations for massless fermions or neutrinos and
give rise to Maxwell's equations, both in momentum space.  In order to
proceed with our constructive program we have now first to imbed two-component
spinor spaces in four component ones  and then in eight
component ones and so on; and at each step analyze which are the
corresponding Clifford algebras  together
with Cartan's equations to see if they may represent some further
elementary laws of physical phenomena.

At first sight this program might appear of difficult realization
since null vectors are ``squares" of spinors and it might
appear difficult to obtain from sums of spinors
sums of vectors necessary for the imbedding of null
spaces in higher dimensional ones. Instead the program is simple
if we adopt Proposition 1 and exploit the properties of simple or
pure spinors.

In fact let us consider $\Cl(2n)= {\rm End} \,S$ and $\psi,\phi \in
S$, then, for $\psi$ arbitrary and $\vf_\pm = 1/2 \left(1 \pm
\g_{2n+1}\right)\f$ simple, the vectors $z^\pm \in W$ with
components
\begin{equation}
    z_a^\pm = \langle B \psi, \g_a \left(1 \pm \g_{2n+1}\right) \f
\rangle ,\qquad a= 1,2,\dots,2n \label{3.1}
\end{equation}
are of the form (2.15) and therefore are null:
\begin{equation}
    z_a^\pm z_\pm^a = 0 .   \label{3.2}
\end{equation}

Let us sum them and we obtain:
\begin{equation}
    z_a^+ + z_a^- = Z_a = \langle B \psi, \g_a \f\rangle \,, \qquad
a = 1, 2,\dots, 2n \,, \label{3.3}
\end{equation}
where $\f = \vf_+ \oplus \vf_- \in S$, while $\varphi_\pm \in S_\pm$
where $\Cl_0(2n) = {\rm End} \,S_\pm$ and therefore:
\begin{equation}
    S = S_+ \oplus S_- \,.\label{3.4}
\end{equation}
This means that starting from two simple spinor spaces we operated
the most obvious operation: their direct sum  spanned by spinors
of double dimension. Let us now examine the corresponding
operation in the vector space $W$; it is represented by eq.(4.3), defining the
$Z_a$, they are, in general, the components of a non null vector of $\C^{2n}$:
\begin{equation}
    Z_a Z^a \not= 0\ . \label{3.5}
\end{equation}
However, $\f =\varphi_+\oplus\varphi_-$ is a $2^n$ component spinor and
as such it could represent a simple spinor of $\Cl_0(2n+2)$. Let
us in fact take for the generators $\Gamma_A$ ($A=1, 2,\dots,
2n+2$) of $\Cl(2n+2)$ the Cartan basis (by which we mean that the
corresponding Dirac spinor is a direct sum of Weyl spinors):
\begin{multline}
\Gamma_A \colon \Gamma_a = \begin{pmatrix}0 & \g_a \cr \g_a & 0 \end{pmatrix} ,
\quad
\Gamma_{2n+1}= \begin{pmatrix} 0 & \g_{2n+1} \cr \g_{2n+1} & 0 \end{pmatrix}
\quad \\ \Gamma_{2n+2}= \begin{pmatrix} 0 & 1_n \cr -1_n & 0 \end{pmatrix}
\label{3.6}
\end{multline}
and the volume element:
\begin{equation}
\Gamma_{2n+3} = \Gamma_1 \Gamma_2 \cdots \Gamma_{2n+2} = \begin{pmatrix}
1_n & 0 \cr 0 & -1_n \end{pmatrix} \label{3.7}
\end{equation}
Then if $\Phi$ is a $2^{n+1}$ component Dirac spinor of $\Cl(2n+2)$ it has,
in the Cartan basis, the form:
\begin{equation}
\Phi= \begin{pmatrix}\f'_+ \cr \f'_- \end{pmatrix} \label{3.8}
\end{equation}
where the $2^n$ component Weyl spinors are given by
\begin{equation}
    \f'_\pm = \frac{1}{2} \left(1 \pm \G_{2n+3}\right) \Phi \label{3.9}
\end{equation}

Suppose they are simple, then the
vectors $Z_\pm \in \C^{2n+2}$ with components:
\begin{equation}
    Z_A^\pm = \langle B \Psi \Gamma_A\left( 1 \pm \Gamma_{2n+3} \right)
\Phi \rangle \label{3.10} ,
\end{equation}
with $\Psi$ arbitrary spinor, are of the form (2.15) and will be null:
\begin{equation}
    {Z_A}^\pm {Z^A}_\pm = 0 \,;\qquad A = 1, 2, \dots, 2n+2
    \label{3.11}
\end{equation}
It is easy to see that in the Cartan-basis (\ref{3.6}) ${Z_A}^\pm$
are expressed by
\begin{eqnarray}
 Z_a^\pm &=& \langle B \psi, \g_a \f'_\pm \rangle\,; \quad a = 1, 2,\dots,
 2n \nonumber \\ Z^\pm_{2n+1} &=& \langle B \psi, \g_{2n+1} \f'_\pm
 \rangle \label{3.12} \\ Z^\pm_{2n+2} &=& \langle B \psi, \pm 1_n
 \f'_\pm \rangle \nonumber
\end{eqnarray}

If we now identify the $2^n$ component simple spinor of $\Cl_0(2n+2)$,
say $\f'_+$, with the Dirac spinor $\f$ of $\Cl(2n)$ appearing in
eq.~(\ref{3.3}) (which is possible because of the isomorphism $\Cl(2n)
\sim \Cl_0(2n+1) \hookrightarrow 2 \Cl_0(2n+1) = \Cl(2n+1) \sim
\Cl_0(2n+2)$) we have that the non null vector $Z \in \C^{2n}$ with
components $Z_a= z^+_a+z^-_a$ given by (\ref{3.3}) is a projection in
$\C^{2n}$ of a null vector of $\C^{2n+2}$ whose two extra components
are
\begin{equation}
  Z_{2n+1} = \langle B \psi, \g_{2n+1} \f \rangle \,; \qquad \quad
  Z_{2n+2}= \langle B \psi, \pm 1_n \f \rangle \label{3.13}
\end{equation}
provided $\f$ may be considered as a simple spinor of
$\Cl_0(2n+2)$, and we have the proposition\\

\noindent\underline{Proposition 2.}\ Given two simple spinors $\vf_\pm$ of $\Cl_0(2n)$ and the
corresponding null vectors $z_\pm \in \C^{2n}$ with components
\begin{equation}
    {z_a}^\pm = \langle B \psi, \g_a \vf_\pm \rangle
    \label{3.14}
\end{equation}
their sum
\begin{equation}
    z_a^+ + z_a^- = Z_a = \langle B \psi, \g_a \f \rangle\,,
    \label{3.15}
\end{equation}
where $\phi=\varphi_+\oplus\varphi_-$, is a projection on $\C^{2n}$ of a null vector
$Z \in
\C^{2n+2}$ which is
obtained by adding to $Z_a$ the two extra components
\begin{equation}
    Z_{2n+1} = \langle B \psi, \g_{2n+1} \f \rangle \qquad  Z_{2n+2} =
\langle B \psi, \pm  \f \rangle \label{3.16}
\end{equation}
provided $\f$ satisfies the conditions of simplicity as a Weyl
spinor of $\Cl_0(2n+2)$.

In this way we see that to the direct sum of simple or pure spinor spaces,
giving rise to a higher dimensional simple spinor space:
\begin{equation}
    S_+ \oplus S_- = S \,, \label{3.17}
\end{equation}
where $\Cl_0(2n) = {\rm End} S_\pm$ and $\Cl(2n+2) = {\rm End} S$, there
correspond the sum of null vectors giving rise to a higher
dimensional null vector:
\begin{equation}
    z^+ +z^- \hookrightarrow  Z  \label{3.18}
\end{equation}
where $z^\pm \in \C^{2n}$ and $Z \in \C^{2n+2}$; and we have an
instrument for imbedding spinor spaces in higher dimensional
spinor spaces and correspondingly null vector-spaces in higher
dimensional null vector spaces.

    The above may be easily restricted to the real spaces, of
interest for physics, in which case the projective null quadric
represent compact manifolds of the type (\ref{1.18}) or
generalizations of them, and then to the direct sums $S_+ \oplus
S_- = S$ of simple spinor spaces there will correspond the
embedding of compact manifolds into compact ones, increasing at
each step the dimension by two. It is interesting to note that
for real spaces the signature of the two extra dimensions will be
uniquely determined by Proposition~2, since $Z_A Z^A=0$ is an
identity when $Z_A$ are bilinearly expressed in terms of spinors
$\psi$ and $\vf$. In particular we adopted it in our
constructive approach of spinor spaces, when starting from two
component spinors we constructed four component spinors and also
the corresponding real, pseudo euclidean vector space, obtained after
substituting $B\psi$ with $B\varphi^c=\varphi^\dagger$ in eq.~(3.4) and
obtaining
eq.~(3.6), with
Lorentzian signature unambiguously defined.

Proposition 2 remains
valid also for $z^\pm$ and $Z$ real with the only difference that
now in eq.(4.16) $Z_{2n+1}$ and $Z_{2n+2}$ have to be substituted with the
real $P_{2n+1} =\langle B\phi^c,\gamma_{2n+1}\phi\rangle$ and
$P_{2n+2} =\langle B\phi^c,\pm i\phi\rangle$, respectively, and the signature of
the $2n+2$ space is unambiguously defined to remain Lorentzian.

This ``construction'' is the simplest and the most natural one since
strictly correlated with the geometry of simple or pure spinors.
Furthermore, it generalizes well established practice with Dirac
spinors: each of them is the direct sum of two Weyl two component
spinors which are simple spinors of four dimensional space-time.
However, the four component Dirac spinor, being direct sum of them, may
only be considered simple insofar isomorphic to a Weyl spinor of
a six dimensional space. Our ``construction'' simply consists in
extending this natural and well known practice to higher
dimensional spaces.

\secs{FROM FOUR TO EIGHT COMPONENT SPINORS}\label{sec4}

    Let us now sum the two null vectors ${p_\mu}^\pm$ of $\R^{3,1}$
defined by eq.~(\ref{2.10}):
\begin{equation}
    p_\mu = p_\mu^+ + p_\mu^- = \tilde \psi \g_\mu \psi , \qquad \qquad
\mu = 0, 1, 2,3 \label{4.1}
\end{equation}
$p_\mu$, because of Proposition 2, are the projection on
$\R^{3,1}$ of a null vector of a six dimensional pseudo-euclidean
space, of definite signature, obtained by adding to the 4 real
components $p_\mu$ the following $p_5$ and $p_6$, both real:
\begin{equation}
    p_5 = \tilde \psi \g_5 \psi \qquad  p_6 = \tilde \psi i \psi
        \tag{\ref{4.1}$'$} \label{4.1p}
\end{equation}
where, as easily verified, $p_a$ ($a = 0, 1, 2, 3, 5, 6$) are real and they
identically satisfy the equation
\begin{equation}
  p_1^2 + p_2^2 + p_3^2 - p_0^2 + p_5^2 + p_6^2 = 0
\label{4.2}
\end{equation}
that is they build up a null vector in $\R^{5,1}$ and their direction
define the projective Klein quadric equivalent to $S^4$. It may be
easily shown that the $p_a$ may be expressed in the following form:
\begin{equation}
    p_a = \Psi^\dagger \Gamma_0 \Gamma_a \left(1 + \Gamma_7\right) \Psi
\label{4.3} \qquad a=1,\dots,6\,.
\end{equation}
where $\Gamma_a$, of the form (4.6), with $\Gamma_6=\sigma_2\otimes 1$, are the generators of $\Cl(5,1)$ (or of
$\Cl(4,2)$ if $\Gamma_6$ is substituted with $i\Gamma_6$), and $\Psi$ an associated eight component spinor.  With
$p_a$ we may generate the Cartan's equation:
\begin{equation}
    \left(p_\mu \g^\mu + p_5 \g_5 +  ip_6 \right) \psi = 0 \label{4.4}
\end{equation}
which, for $\psi$ Majorana, since $p_6 \equiv 0$, reduces to:
\begin{equation}
    \left(p_\mu \g^\mu + \g_5 p_5\right) \psi = 0 \label{4.5}
\end{equation}
where
\begin{equation}
    p_\mu p^\mu = - p_5^2   \tag{\ref{4.5}$'$} \label{4.5p}
\end{equation}
and may therefore give origin to the Dirac equation in momentum space
$\R^{3,1}$.  Obviously there will be another such equation for the
projector $\left(1-\Gamma_7\right)$.

Observe that eq.~(5.5) may be also directly obtained from the
extension of Cartan's eq.~\eqref{1.1} for Dirac spinors associated
with $\Cl(3,1)$ to spinors associated with $\Cl_0(4,1)$; which is
allowed due to isomorphism of simple algebras (2.7$'$). As such
$\psi$ is a Pauli spinor.

We may now also consider the Clifford algebra $\Cl(4,2) = {\rm End}
S$ corresponding to $\R^{4,2}$, the conformal extension of space-time, with
generators
\begin{equation}
    \G_0, \G_1, \G_2, \G_3, \G_5, \G_6  \label{4.6}
\end{equation}
and volume element $\G_7 = -i \G_0 \G_1 \G_2 \G_3 \G_5 \G_6$. Its simple
Weyl spinors $\pi_+, \pi_-$, defined by
\begin{equation}
    \pi_\pm = \frac{1}{2}\left( 1 \pm \G_7\right) \Psi \label{4.7}
\end{equation}
where $\Psi \in S$, are vectors of spinor spaces $S_\pm$ such that
$\Cl_0(4,2) = {\rm End} S_\pm$, and obey the Cartan's equation
\begin{equation}
     z_a^\pm \G^a\left(1 \pm \G_7\right) \Psi = 0\qquad
a=1,2,\dots ,6 \label{4.8}
\end{equation}
where
\begin{equation}
     {z_a}^\pm = \frac{1}{4} \langle B \Psi^c, \G_a \frac{1}{2} \left(1
\pm \G_7\right) \Psi \rangle \label{4.9}
\end{equation}
and $B\Psi^c = \Psi^\dagger \G_0 \G_6$.

Because of Proposition 1 they are complex ($n = 3$) and precisely
\begin{equation}
    {z_a}^+ = - {{{\bar z}_a}}^- \label{4.10}
\end{equation}
and therefore eq.~(\ref{4.8}) may not have immediate physical
interpretation as real vectors in momentum space; they could
instead represent charged currents and as such explain the
geometrical origin of the electro-weak model as it will be shown
in section 6.2. The Weyl simple spinors $\pi_\pm$ defined by
(\ref{4.7}), associated with $\R^{4,2}$ were named twistors by R.
Penrose~\cite{9}, and have had several interesting applications
in mathematics and general relativity.

\secs{EIGHT COMPONENT SPINORS} \label{sec5}

\subsection{THE NUCLEON DOUBLET} \label{sec5.1}

Let us consider the real null vectors given by the generalization
of eq.~(\ref{4.3}):
\begin{equation*}
 {p_a}^\pm =  N^\dagger \G_0 \G_a \left(1 \pm\G_7\right) N
 \qquad a=1,\dots,6\,.
\end{equation*}
where $N$ is an eight component Dirac spinor of $\Cl(5,1)$ and let
us sum them:
\begin{equation}
 \frac{1}{2}\left({p_a}^+ + {p_a}^-\right) = p_a =
 N^\dagger \G_0 \G_a N .    \label{5.1}
\end{equation}

    Then, because of Proposition 2, together with the components $p_7 =
N^\dagger \G_0 \G_7 N$ and $p_8 =  N^\dagger \G_0 iN$ they build
up, for $N$ simple spinor of $\Cl_0(7,1)$, a real null vector $P
\in \R^{7,1}$ with components:
\begin{equation}
  p_A = N^\dagger \G_0 \G_A N \quad {\rm with} \quad \G_A =
  \left\{ \G_a, \G_7 ,-i\right\} \label{5.2} \qquad A=1,\dots,8\,.
\end{equation}
In fact if the spinor $N$ is in the Dirac basis (by which we mean that $N$ is
the direct sum of two Dirac spinors which is allowed because of the
isomorphisms \eqref{1.7} and (2.7$'$))
\begin{equation*}
N = \begin{pmatrix} \psi_1 \\ \psi_2 \end{pmatrix}
\end{equation*}
where $\psi_j$ are Dirac spinors of $\Cl(3,1)$, $\Gamma_A$ have
the form:\footnote{The spinors $\psi_1,\psi_2$ in the $N$ doublet
may be either Weyl spinors of the automorphism space of
$\Cl_0(5,1)$, or Pauli of $\Cl(4,1)$ or Dirac of $\Cl(3,1)$.
Correspondingly, either $\Gamma_\mu = \sigma_1\otimes\gamma_\mu$
or $\sigma_2\otimes\gamma_\mu$ for Weyl, or $\Gamma_\mu =
\sigma_3\otimes\gamma_\mu$ for Pauli, or $\Gamma_\mu =
1\otimes\gamma_\mu$ for Dirac. These are then $3+1$ independent
representations corresponding to the four dimensional space of
quaternions.}
$$\G_\mu = \begin{pmatrix}\g_\mu & 0 \cr 0 & \g_\mu \end{pmatrix} ,
  \quad \G_5 = \begin{pmatrix}0 & \g_5 \cr \g_5 & 0 \end{pmatrix} ,
  \quad \G_6 = \begin{pmatrix} 0 & -i\g_5 \cr i\g_5 & 0 \end{pmatrix}, $$
\begin{equation}
 \G_7 = \begin{pmatrix}\g_5 & 0 \cr 0 & -\g_5 \end{pmatrix} \label{5.6}
\end{equation}
and it may be easily verified that, in accordance with Proposition 2,
$p_A$ given by (6.2) satisfies to:
\begin{equation}
    p_\mu p^\mu + p_5^2 + p_6^2 + p_7^2 + p_8^2 = 0 . \label{5.4}
\end{equation}

The Cartan's equation for $N$, taking into account of (6.2) and of (6.3) is:
\begin{equation}
\left( p_\mu \cdot 1 \otimes \g^\mu + {\vec \pi} \cdot {\vec \s} \otimes
\g_5 + iM\right) N = 0 %%%%%%%\tag{\ref{6.5}}\label{5.5p}
\end{equation}
where, according to eq.~(6.2):
\begin{equation}
p_\mu =\frac{1}{8}(\tilde\psi_1\gamma_\mu\psi_1+\tilde\psi_2\gamma_\mu\psi_2);\
\ {\vec \pi} = \frac{1}{8}{\tilde N} {\vec \s}\otimes \g_5 N ; \ \
 M = \frac{i}{8} \left({\tilde \psi}_1 \psi_1 + {\tilde \psi}_2
 \psi_2\right) \label{5.7}
\end{equation}
where:
\begin{equation*}
 {\tilde N} = \begin{pmatrix} {\tilde \psi}_1 & {\tilde \psi}_2 \end{pmatrix}.
\end{equation*}
Eq.~(6.5) represents the well-known proton-neutron equation
when interacting with the pseudo-scalar pion isovector in momentum
space, and with (6.6), eq.~(6.5) is an
identity. Observe that the pseudoscalar nature of the pion
derives from imposition that $N$ is a doublet of Dirac spinors
which, in turn, imposes the representation (6.3) of $\G_a$ where
$\g_5$ must be contained in $\G_5$, $\G_6$, $\G_7$ (in order to
anticommute with $\G_\mu$). Again the pion field $\vec{\pi}$
appears here as bilinearly expressed in terms of the
proton-neutron field. However this does not imply that in the
quantized theory  the pion should be considered as a bound state
of proton-neutron.

Observe that from eq.~(6.6) we have: $p_\mu = \frac{1}{8}
\left({\tilde \psi}_1 \g_\mu {\psi}_1 + {\tilde \psi}_2 \g_\mu
{\psi}_2\right)$ which could then be interpreted as total
momentum of the nucleon doublet (in absence of pions) and as such
represent the translation operator giving rise to Poincar\'e translations and to space-time $\R^{3,1}$, while $p_5$,
$p_6$, $p_7$ are extra momenta which represent interaction terms
and, in principle, do not need the corresponding dimensions in configuration space. Momentum space is here compact and represented by the projective quadric
(\ref{5.4}); that is $S^6/{\Z}_2$, whose radius, in the physical
interpretation may be identified with a mass. We will see in
Chapter 13 that in eq.~(6.5) the mass term $M$ has to be set to
zero if to $N$ the simplicity or pureness constraint is imposed.

The term ${\vec \pi} \cdot {\vec \s} \otimes \g_5$ presents the
so-called isospin internal symmetry  $SU(2)$ of the nucleon
doublet which, in view of the above geometrical derivation
deserves further comments. We will deal with them in
section~\ref{sec6}.

\subsection{THE ELECTROWEAK MODEL} \label{sec5.2}

With eight component spinors we may bilinearly obtain two more real null
vectors in eight dimensional spaces. In fact as a consequence of
Corollary 1, for $n= 4$, the vectors $K^\pm \in \R^{5,3}$ are real and
null. Its real components are [4]:
\begin{equation}
  {K_A}^\pm = \langle \Theta^\dagger G_0 G_6 G_8 G_A \left(1 \pm G_9\right)
  \Theta \rangle,   \qquad A = 0, 1, 2, 3, 5, 6, 7, 8 \label{5.8}
\end{equation}
where $G_A$ are the generators of $\Cl(5,3) ={\rm End} S$, $\Theta\in S$ and we suppose
that the eight component spinors:
\begin{equation}
    \Psi_\pm = \frac{1}{2} \left(1 \pm G_9\right) \Theta \label{5.9}
\end{equation}
are simple Weyl spinors (subject to one constraint equation).

If we take $G_A$ in the Cartan basis, eq. (\ref{5.8}) may be
expressed in the form:
\begin{equation}
{K_A}^\pm = -\frac{1}{8} \Psi_\pm^\dagger \G_0 \G_6 \G_7 \G_A \Psi_\pm
\label{5.10}
\end{equation}
where
\begin{equation*}
  \G_A=\left\{ \G_a, \G_7, \pm
  1\right\},
\end{equation*}
and $\G_a$ are generators of $\Cl(4,2)$.

The corresponding Cartan's equations are:
\begin{equation}
  {K_A}^\pm G_A \left(1 \pm G_9\right) \Theta = 0 \,.   \label{5.11}
\end{equation}
These equations, as well as equation (6.5),
might also have physical meaning.  To obtain it let us first observe
that the components $K_A^\pm$ of the null vector $K^\pm \in \R^{5,3}$
may be also obtained from Proposition 2. In fact let us start from the
six dimensional complex null vectors with components ${z_a}^\pm$ given
by (\ref{4.9}) if we subtract them:
\begin{equation}
  {z_a}^+ - {z_a}^- = {z_a}^+ + {{{\bar z}_a}}^+ = K_a \label{5.12}
\end{equation}
we obtain a real six component vector of $\R^{4,2}$ which identifies
with the first six components of $K_A$ given by eq.~(\ref{5.10}), in
fact it is easily seen that:
\begin{equation}
K_a = - \frac{1}{8} \Psi^\dagger \G_0 \G_6 \G_7 \G_a \Psi \qquad
a=0,1,2,3,5,6 \label{5.13}
\end{equation}
that is the first six components of ${K_A}^\pm$, for $\Psi=\Psi_+$,
say: simple spinor of $\Cl(5,3)$.

The main difference between the null vectors $p_A$ of the previous
section and the ${K_A}$ of this one is that while $p_a$ are sums of
real vectors (see eq.~(6.1)) bilinear in Dirac spinor, the
$K_A$ are sums of complex vectors (see eq.~(6.11)), bilinear in
twistors. Therefore their four component space-time vectors may
represent neutral and charged currents respectively which may
constitute the basic ingredient of the Salam-Weinberg electroweak
model, as we will see.

Let us first write the Cartan's equation with $K_A$:
\begin{equation}
\left( K_\mu \G^\mu + K_5 \G^5 - K_6 \G^6 + K_7 \G^7 + K_8 \right)
\begin{pmatrix} \pi_+ \cr \pi_- \end{pmatrix} =0 \label{5.14}
\end{equation}
where we set $\Psi = \begin{pmatrix}\pi_+ \cr \pi_- \end{pmatrix}$ and by
construction $\pi_\pm$ are Weyl spinors of $\Cl_0(4,2)$ or twistors, and
therefore the $\G_a$ are in the Cartan basis (\ref{3.6}) and we may write
eq.~(\ref{5.14}) in matrix form as follows:
\begin{equation}
  \begin{pmatrix} K_7 + K_8 & K_\mu \g^\mu + K_5 \g_5 + K_6 \cr K_\mu
  \g^\mu + K_5 \g_5 - K_6 & - K_7 + K_8 \end{pmatrix} \begin{pmatrix}
  \pi_+ \cr \pi_- \end{pmatrix} =0  \tag{\ref{5.14}$'$} \label{5.14p}
\end{equation}
Observe that the first six components of $K_A$ are given by (\ref{4.9})
where:
\begin{equation*}
{z_a}^\pm = \frac{1}{4} \Psi^\dagger \G_0 \G_6 \G_a \frac{1}{2} \left( 1
\pm \G_7 \right) \Psi \in M\left( \pi_\pm \right)
\end{equation*}
and $\pi_\pm$ obey the Cartan's equations:
\begin{equation}
  \left( z_\mu^+ \g^\mu + z^+_5 \g_5 - z_6^+ \right) \pi_+ =0 \qquad
  \left( {\bar z}_\mu^+ \g^\mu + {\bar z}^+_5 \g_5 + {\bar z}_6^+
  \right) \pi_- =0 \,.    \label{5.15}
\end{equation}
Therefore expressing the $K_a$ in (6.13$'$)) as ${z_a}^+ + {{\bar
z}_a}^+$ and taking into account of (\ref{5.15}) we arrive at:
\begin{equation}
  \begin{pmatrix}\frac{i}{4}{\tilde \pi}_+ \pi_+ & \frac{1}{2}\left(
  z_\mu \g^\mu +z_5\g_5+z_6\right) \cr \frac{1}{2}\left( {\bar z}_\mu
  \g^\mu +{\bar z}_5\g_5-{\bar z}_6\right) & \frac{i}{4}{\tilde \pi}_-
  \pi_- \end{pmatrix} \begin{pmatrix}\pi_+ \cr \pi_- \end{pmatrix} =0
  \label{5.14pp} \tag{\ref{5.14}$''$}
\end{equation}
where the $+$ superscript for $z_a$ was suppressed and it was
taken into account that
\begin{equation}
K_7 =\frac{i}{8}\left( {\tilde \pi}_+ \pi_+ - {\tilde \pi}_- \pi_-\right)
\qquad
K_8 =\frac{i}{8}\left( {\tilde \pi}_+ \pi_+ + {\tilde \pi}_- \pi_-\right)
\label{5.16}
\end{equation}
Let us express the complex six-vector $z_a$ in the form
\begin{equation}
z_\mu = A^{(1)}_\mu - i A^{(2)}_\mu , \quad z_5= \pi_1 - i \pi_2, \quad
z_6=s_2-i s_1 \label{5.17}
\end{equation}
where $A^{(i)}_\mu$ , $\pi_i$, $s_i$ are real. Then taking into account
that
\begin{equation}
    z_a = \frac{1}{4} {\tilde \pi}_+ \g_a \pi_- \label{5.18}
\end{equation}
we easily arrive at:
\begin{eqnarray}
A^{(j)}_\mu &=& \frac{1}{8} {\tilde \Psi} \s_j \otimes \g_\mu\Psi \nonumber \\
\pi_j &=& \frac{1}{8} {\tilde \Psi} \s_j \otimes \g_5\Psi \qquad \quad j =
1, 2 \label{5.19} \\
s_j &=& \frac{1}{8} {\tilde \Psi} \s_j \otimes 1 \Psi \nonumber
\end{eqnarray}
where ${\tilde \Psi}= \begin{pmatrix} {\pi_+}^\dagger \g_0 &
{\pi_-}^\dagger \g_0 \end{pmatrix}$.

Setting these in eq.~(6.13$''$)) we easily arrive at:
\begin{equation}
\left\{ \sum\limits_{j=1}^2 \left( A_\mu^j \s_j \otimes \g^\mu + \pi^j \s_j
\otimes \g_5 + s^j \s_j \otimes 1 \right) + M \right\} \begin{pmatrix}
\pi_+ \cr \pi_- \end{pmatrix}=0
\tag{\ref{5.14}$'''$}\label{5.14ppp}
\end{equation}
where
\begin{equation}
  M=\begin{pmatrix} \frac{i}{4} {\tilde \pi}_+ \pi_+ & 0 \cr 0 & \frac{i}{4}
  {\tilde \pi}_- \pi_- \end{pmatrix} \label{5.20}
\end{equation}
We can now go back to the real null vector $p_A$ defined by
(6.1) which defines the Cartan's equation (6.5) for the spinor
$N = \begin{pmatrix}\psi_1 \cr \psi_2 \end{pmatrix}$, where now $\psi_1$,
$\psi_2$ are Dirac spinor of $\Cl(3,1)$. Following the same procedure as
above we arrive at:
\begin{equation}
\left[ \left( p_\mu 1 + A_\mu^3 \s_3 \right) \otimes \g^\mu + {\vec \pi}
\cdot {\vec \s} \otimes \g_5 + M_1 \right] N =0 \label{5.21}
\end{equation}
where
\begin{equation*}
p_\mu = \frac{1}{16} \left( {\tilde\psi}_1 \g_\mu \psi_1 + {\tilde\psi}_2
\g_\mu \psi_2 \right) \qquad A_\mu^3 = \frac{1}{8} {\tilde N} \s_3 \otimes
\g_\mu N
\end{equation*}
and $M_1$ is given by eq.(\ref{5.20}) where $\psi_1$, $\psi_2$
substitute $\pi_+$, $\pi_-$.

Now observe that we obtained eq.(6.5), for $N$ in the frame of our
programme of constructing fermion multiplets, that is multiplets
of $\Cl(3,1)$ - Dirac spinors, where we exploited the isomorphism
$\Cl_0(2n+2)\simeq \Cl(2n+1)$ and consequently obtained
$\Psi_D=\psi_D\oplus\psi_D$. With a similar construction we could
also have obtained eq.(6.13) for $\Psi =\begin{pmatrix} \pi_+\\
\pi_-\end{pmatrix}$; we had only to ignore that isomorphism and
exploit just the last step of the construction:
$\psi_W\oplus\psi_W=\Psi_D$, which means that $N$ and
$\Psi$ may be correlated; and in fact define the chiral
projectors:
\begin{equation}
L=\frac{1}{2} (1+\g_5);\quad R=\frac{1}{2} (1-\g_5)
\end{equation}
We have that the unitary matrix
\begin{equation}
U=\left\vert \begin{matrix} L&R\\ R&L \end{matrix}\right\vert
=U^{-1}
\end{equation}
transforms $\Psi$ in $N$ and vice versa:
\begin{equation}
UN=\left\vert \begin{matrix} L\psi_1+R\psi_2\\ R\psi_1+L\psi_2
\end{matrix} \right\vert = \Psi = \left\vert \begin{matrix}
\pi_+\\ \pi_- \end{matrix}\right\vert
\end{equation}

We will indicate with $\psi_{iL}=L\psi_i$ and with
$\psi_{iR}=R\psi_i$ the left-handed and right-handed parts of
$\psi_i,(i=1,2)$ and the same for $\pi_\pm$. We have then from
(6.28):
$$
\Psi =\begin{pmatrix} \pi_{+L}\\ \pi_{+R}\\ \pi_{-L}\\ \pi_{-R}
\end{pmatrix} = \begin{pmatrix} \psi_{1L}\\ \psi_{2R}\\
\psi_{2L}\\ \psi_{2R}\end{pmatrix}
$$
which means (since $LR=0$) that
\begin{equation}
L\Psi =\Psi_L=LN=N_L
\end{equation}
That is, the left-handed parts of $\Psi$ and of $N$ are identical.

Suppose now that $\psi_1=e$ and $\psi_2=\nu_L$ represent an
electron and a left-handed massless neutrino respectively. Then
let us act with $R$ on eq. (6.13$'''$) and sum it to (6.20) and,
remembering that $R\g_\mu =\g_\mu L$, and that $N=N_L+N_R$ we
obtain (after eliminating the pion interaction terms):
\begin{equation}
(p_\mu\g^\mu+im)\begin{pmatrix} e\\ \nu\end{pmatrix} +\vec
A_\mu\cdot \vec\s\otimes\g^\mu\begin{pmatrix} e_L\\
\nu_L\end{pmatrix} +(B_\mu\g^\mu +\tau )e_R =0
\end{equation}
where $\tau =s_1-is_2$. $\vec A_\mu$ is an isotriplet and $B_\mu$
a singlet. This equation derives from a Lagrangian which, together
with one for the vector fields $\vec A_\mu$ and $B_\mu$, may be
assumed at the basis of the electroweak model.

Observe that the $A^{(3)}_\mu$ component of $\vec A_\mu$ and
$B_\mu$ representing neutral vectors is built up from the real
vector $p^{(1)}_\mu$ and $p^{(2)}_\mu$ while $A^{(1)}_\mu
+iA^{(2)}_\mu =z_\mu$ representing changed vectors responsible of
weak interactions are bilinearly constructed with twistors, which
might have consequences of interest.

It may be shown~\cite{9} that if the triplet $\left(e_L, e_R, \nu_L
\right)\equiv \left( e_L, e_L^c, \nu_L\right)$, where $e^c$ means
charge conjugate of $e$, is subject to an $SU(3)$ symmetry, then the
mixing angle $\theta$ of the model is fixed such that $\sin^2 \theta =
0.25$.

\subsection{THE NEUTRAL-CHARGED FERMION DOUBLETS} \label{sec5.3}

The possible geometrical origin of isospin symmetry $SU(2)$ seen
above, might also naturally explain the frequent appearance, in
the elementary particle landscape, of the charged-neutral fermion
doublets like proton-neutron, electron-neutrino, muon-neutrino
etc.

 Let us in fact write down explicitly eq.~(6.5) in terms of the
 two Dirac spinors $\psi_1$ and $\psi_2$ of the $N$ doublet:
%\begin{equation}
$$
\begin{array}{rl}
     \left( p_\mu \gamma_\mu + p_7 \gamma_5 + ip_8 \right) \psi_1 +
     \gamma_5 \left( p_5 - i p_6 \right) \psi_2 &= 0\, ,  \\
     \left( p_\mu \gamma_\mu - p_7 \gamma_5 + ip_8 \right) \psi_2 +
     \gamma_5 \left( p_5 + i p_6 \right) \psi_1 &= 0\, .
  \end{array}
\eqno(6.5')$$
%  \label{5.5pp} \tag{\ref{6.5}$'$}
%\end{equation}
Now all $p_a$ are real, therefore defining
\begin{equation}
   p_5 \pm i p_6 = \rho e^{\pm i \frac{\omega}{2}}
\end{equation}
and multiplying the first eq.~(6.5$'$) by $e^{i \frac{\omega}{2}}$
we obtain
\begin{equation}
\begin{split}
  \left( p_\mu \gamma_\mu + p_7 \gamma_5 + ip_8 \right) e^{ i
  \frac{\omega}{2}} \psi_1 + \gamma_5 \rho \psi_2 &= 0\,, \\ \left(
  p_\mu \gamma_\mu - p_7 \gamma_5 + ip_8 \right) \psi_2 + \gamma_5 \rho
  e^{i \frac{\omega}{2}} \psi_1 &= 0\,.
\end{split}
\label{5.32}
\end{equation}
We see then that $\psi_1$ appears with a phase factor $e^{ i
\frac{\omega}{2}}$ corresponding to a rotation through an angle
$\omega$ in the circle defined by
\begin{equation}
  p_5^2 + p_6^2 = \rho^2 \label{5.33}
\end{equation}
in the vector space of the Klein quadric defined by
eq.~\eqref{5.4}, which in turn corresponds to an imaginary
dilation in $\R^{4,2}$. In fact the corresponding transformation
in spinor space is generated by the Lie algebra element
\begin{equation}
  J_{56} = \frac{1}{2} \left[ \G_5,  \G_6 \right]. \label{5.34}
\end{equation}
which is obtained from $J_{56}$ in the $SU(2,2)$ covering the
conformal group, after multiplying the generation $\Gamma_6$ by
the imaginary unit $i$.

Observe that this complexification was intrinsic to the modality
of our construction which brought us to $\Cl(5,1)$ which may be
obtained by setting an imaginary unit factor $i$ in front of the
generator $\G_6$ of $\Cl(4,2)$ and which finally generated the
$SU(2)$ internal symmetry. In this way $i\Gamma_6$ may be
interpreted as a generator of reflections with respect to the sixth
(time-like) axis (see Chapter 7) for $\Cl(4,2)$.

With this interpretation the dilation covariance of the
complexified conformal group induced the $U(1)$ group of symmetry
represented by the phase factor in front of $\psi_1$ and not of
$\psi_2$ (or vice-versa).

If we now translate this in the corresponding Fourier dual Minkowski
space-time, since dilation
covariance is local, we may consider the phase angle $\omega$ as
coordinate dependent $\omega \rightarrow \omega\left(x\right)$ and
then to maintain the covariance of eq.~\eqref{5.32} we will have
to introduce an abelian gauge potential $A_\mu$ interacting with
$\psi_1$ only and, as easily seen, the eq.~(6.27) will
become, in space-time $\R^{3,1}$:
\begin{equation}
  \left\{ \gamma_\mu \left[ i \frac{\partial}{\partial x_\mu} +
  \frac{e}{2} \left( 1 -i \G_5 \G_6 \right) A_\mu \right] + \vec{\pi}
  \cdot \vec{\sigma} \otimes \gamma_5 + M \right\} \begin{pmatrix} P
  \cr N \ \end{pmatrix} = 0 \label{5.35}
\end{equation}
well representing the equation of the proton-neutron doublet
interacting with the pion and with the electromagnetic potential
$A_\mu$.

Observe that the electric charge $Q$ of the nucleon doublet $N$ is
then represented by:
\begin{equation}
  Q = \frac{e}{2} \left( 1-i \G_5 \G_6 \right), \label{5.36}
\end{equation}
where $-i\G_5\G_6=\sigma_3 \otimes 1$, that is the third component
of isospin generator which, with the other two components $\s_1
\otimes 1$ and $\s_2 \otimes 1$ generate $SU(2)$ isospin symmetry
of nuclear forces.

It is interesting to observe that, in the frame of the study of
conformal covariance of spinor field theory, the existence of non
equivalent spinor structure for $\Cl(4,2)$ was pointed out by
L.~Dabrowski~\cite{Ludwik} and correlated with exotic spinors studied
by Petri~\cite{Petri}. Here we see that they may be directly derived
from Cartan's equations if interpreted as equations of motion in
momentum space and that the local phase factor of the charged partner
of the spinor doublet may be interpreted as the result of a
complexified dilation.

As we will see this might have interesting consequences also in higher
dimensional spaces.

Concluding this chapter we have seen that the minimal Clifford
algebras to accommodate eight component spinors are the ones of
eight dimensional pseudo-euclidean spaces $\R^{5,3}$ or $\R^{7,1}$
and they give rise to equations (6.5), (6.25) and (6.30) of
physical significance in momentum space.

    Several more aspects should be further analyzed, among these
the simplicity constraint for eight component simple or pure
spinors and the role of triality, discovered by Cartan, which
poses on equal footing the eight component spinors $\frac{1}{2} (1
\pm G_9) \Theta$ and the null vectors, they will be outlined in
Chapter 13. Furthermore it may be expected that octonion field of
numbers might play a role, as shown in ref.~\cite{10}. In
particular in the above derivation of eq.~(6.25) the neutral
vector $A^3_\mu$ given in eq.(\ref{5.21}) identifies with equation
representing neutral pion $\pi_3$ in eq.(\ref{5.6}) by
substituting $\gamma_\mu$ with $\gamma_5$. That is
\begin{equation*}
 \pi_3 = A^3_5 .
\end{equation*}
This could represent the starting point for an attempt of the
unification of electroweak and strong interactions in a de Sitter
(or anti-de-Sitter) symmetric theory, as will be discussed
elsewhere.

\secs{REFLECTIONS}\label{sec6}

%\subsection{CONFORMAL REFLECTIONS} \label{sec6a}

    It is well known that if $\g_a$ are the generators of a Clifford
algebra $\Cl(m,n) ={\rm End} S$ corresponding to a
pseudo-euclidean space $V = \R^{m,n}$ the commutators $S_{ab} =
1/2 \left[\g_a,\g_b\right]$ are the elements of the Lie algebra
of spin $(m,n)$, the covering group of $SO(m,n)$ acting on $S$,
while the $\g_a$ are operators on $S$ corresponding to
reflections with respect to an hyper-plane orthogonal to the unit
vector $u_a \in V$. Precisely for any $\psi \in S$ the mentioned
reflections induce the transformation~\cite{1}:
\begin{equation}
    \psi \rightarrow \psi' = \pm \g_a \psi. \label{6.1}
\end{equation}
Since obviously the square of a reflection must be equal to the
identity one must have:
\begin{equation}
    \g_a^2 = 1. \label{6.2}
\end{equation}
Therefore if $\g_a$ represents a time-like unit vector, the
corresponding reflection must be represented by $i \g_a$.

In Dirac's as well as in Cartan's equations like eq.~(\ref{1.1}) the
$\g_a$ appearing in them represent then reflection operators in spinor
space.

Let us now consider the space-time conformal group. As well known
it may be linearly represented by $SO(4,2)$ acting on $\R^{4,2}$
with Clifford algebra $\Cl(4,2)$ whose generators $\G_a$ are
given by (6.3), where $\Gamma_6$ should be substituted by
$i\Gamma_6$ (in the Dirac basis), together with the volume element
$\G_7$, which extends the Clifford algebra to $\Cl(5,2)$. The
reflection operators with respect to the corresponding $5^{th}$,
$6^{th}$ and $7^{th}$ hyper-plane will then be represented, in
spinor space, by:
\begin{equation}
\G_5, i \G_6, \G_7 \label{6.3}
\end{equation}
satisfying the condition (\ref{6.2}); and these are precisely the
operators which appear in eq.~(6.5) which give rise to the
term ${\vec \pi} \cdot {\vec \s} \otimes \g_5$
which, formally, is identical to the one
traditionally introduced to represent the isospin symmetry
$SU(2)$ of pion-nucleon interaction, which then could be ascribed
to the reflection operators~(\ref{6.3}), which are reflections of
the conformal group (and its volume element). Observe that, when
acting in the space of the spinor doublet $N$, they may be
correlated with the quaternion field of numbers (clearly visible
for $\Cl(1,7)$; see section 11.2, eq.~(11.5)).

That the conformal group might be at the origin of isospin symmetry
$SU(2)$ is an old idea, based on the fact that in a certain
representation (the Dirac one) the eight component spinor associated
with $\Cl(4,2)$ is represented by a doublet of Dirac spinors, to be
identified with the proton-neutron doublet. In fact the idea was first
conjectured by W.  Heisenberg, the discoverer of isospin. However the
obstacle to overcome was twofold. First imbedding $SO(3,1)$ in
$SO(4,2)$ or, better its covering $SL(2,\C)$ in $SU(2,2)$ one obtains
$SU(1,1)$ and not $SU(2)$. Second the well-known O'Raifeartaigh no-go
theorem. In our derivation of the traditional equation (6.5)
representing $SU(2)$ isospin symmetry, from spinor geometry, both
obstacles are avoided since $SU(2)$ derives from a reflection algebra
and because of this the no-go theorem does not apply.  Furthermore, as
we have seen from spinor geometry, we naturally derive, through
Proposition 2, the real vector $p_A$ given by eq.~(\ref{5.2})
which is null in $\R^{7,1}$ as shown in
eq.~(\ref{5.4}).

The fact that the conformal reflections, represented by
(\ref{6.3}), might have an important role, should be welcome
to those, like us, who think that the conformal group is a fundamental
group of nature, like Maxwell's equation conformal covariance seems to
suggest; since otherwise we would have a group in which the first four
(space and time) reflections play an important role (parity,
antimatter) while the last two (or three) do not.

The validity of conformal symmetry (for massless systems) has
induced several authors to conjecture, that Minkowski space-time
$M$ may be densely contained in conformally compactified space
time $M_c = \left( S_3 \times S_1 \right) /\Z_2$. There are good
arguments~\cite{12} to think that, in this case, also the Fourier
dual momentum space $P$ should be densely contained in
conformally compactified momentum space $P_c = \left( S_3 \times
S_1 \right)/\Z_2$. $M_c$ and $P_c$ build up conformally
compactified phase space where both concepts of infinite and
infinitesimal ( both infrared and ultraviolet divergences ) would
be {\it a priori} absent. In this frame conformal reflections
could play an important role, of relevance for physics, since
they map $M_c$ to $P_c$ and therefore space time
to momentum space and vice-versa. And since, as we have shown, momentum space is
the appropriate arena for the description of the equations of
motion of fermions in first quantization, in spinor geometrical
form, while space time is appropriate for the description of
classical mechanics in euclidean geometrical form, conformal
reflections could help to understand the meaning of the
correspondence principle as discussed in ref.~\cite{12}.

\secs{SIXTEEN COMPONENT SPI\-N\-ORS} \label{sec8}

\subsection{THE BARYON-LEPTON DOUBLETS} \label{sec:8.1}

 The construction may be continued observing that the real null vector
with components $p_A$ given by eq.~\eqref{5.2} for $N$, thought
as a Weyl spinor of $\Cl_0(7,1)$, may be considered as a
particular case of of the following equation:
\begin{equation}
  P_A^\pm = \Theta^\dagger G_0 G_A \left( 1 \pm G_9 \right)
  \Theta\,, \qquad A=1,2,\dots,8 \,, \label{8.1}
\end{equation}
where $\Theta$ is a sixteen component spinor associated with
$\Cl(7,1)$ of which $G_A$ are the generators and $G_9$  the
volume element.

 Again $\Theta$ may be considered in the Dirac basis
\begin{equation}
  \Theta = \begin{pmatrix} N_1 \cr N_2 \end{pmatrix}
  \label{8.2}
\end{equation}
where $N_1$ and $N_2$ are Dirac spinors of $\Cl(5,1)$ and, again,
defining
\begin{equation}
  P_A = P_A^+ + P_A^- = \Theta^\dagger G_0 G_A \Theta \qquad
  A=1,2,\dots,8 \label{8.3}
\end{equation}
and
\begin{equation}
\begin{split}
  P_9 &= \Theta^\dagger G_0 G_9 \Theta \\
  P_{10} &= \Theta^\dagger iG_0 \Theta
\end{split}
\label{8.4}
\end{equation}
we obtain the components $P_\alpha$ of a ten dimensional real null
vector which defines the Cartan's equation for the spinor $\Theta$:
\begin{equation}
  \left( P_\mu G^\mu + P_5 G_5 + P_6 G_6 + P_7 G_7 + P_8 G_8
  + P_9 G_9 + i P_{10} \right) \Theta =0 \,,
\label{8.5}
\end{equation}
where $G_\alpha$ are the generators of $\Cl(8,1)$. For $\Theta$ in
the Dirac basis we may assign to the $G_\alpha$ the following
form:
\begin{equation}
\begin{split}
 G_\alpha = \left\{ G_a = \begin{pmatrix} \G_a & 0 \cr 0 & \G_a
 \end{pmatrix} ; \, G_7 = \begin{pmatrix} 0 & \G_7 \cr \G_7 & 0
 \end{pmatrix} ;\,  \right. \mspace{35mu} \\  \left.
  G_8  = \begin{pmatrix} 0 & - i\G_7 \cr i\G_7 & 0
 \end{pmatrix} ;  G_9 = \begin{pmatrix} \G_7 & 0 \cr 0 & - \G_7
 \end{pmatrix} \right\} \ \ a=1,2,\dots ,6.
\end{split}
\label{8.6}
\end{equation}
With the same procedure as that of section 6.3, it is easy to show that
the Dirac spinors $N_1$ and $N_2$ obey a system of equations where
$N_1$ (or $N_2$) is multiplied by a phase factor $e^{i \tau/2}$, where
the angle $\tau$ represents a rotation in the circle
\begin{equation}
   P_7^2 + P_8^2 = \rho^2 \label{8.7}
\end{equation}
for which eq.~\eqref{8.5} is covariant ( in spinor space it is
generated by $G_7G_8$). This $U(1)$ symmetry of $N_1$ (or $N_2$)
may be interpreted as a charge which, being different from the
electric charge (generated by $\G_5\G_6$), could be the charge of
strong forces. In his case then, $N_1 = \begin{pmatrix} \psi_1 \cr
\psi_2 \end{pmatrix} $ could represent the nucleon doublet while
$N_2 = \begin{pmatrix} \psi_3 \cr \psi_4 \end{pmatrix} $ the
electron neutrino doublet say, both of which contain an
electrically charged and neutral component.

We could consider then the strong charge as an
eigenvalue of $-iG_7 G_8= \s_3 \otimes \one_8$, while the electric
charge as an eigenvalue of $-i\G_5\G_6=\s_3\otimes \one_4$. There
is a similarity with Dirac four component spinor $\psi$ conceived
as doublet of Weyl $\vf_+$, $\vf_-$ spinors studied in
chapter~\ref{sec2}. There $\g_5 = \s_3 \otimes 1_2$ represents
chirality, while $\s_3$ the third component intrinsic angular
momentum. As we have seen from eq.~\eqref{2.11}, this component
equals $\pm \tfrac{1}{2} \hbar$ on $\vf_\pm$. The corresponding
equation for the quadruplet of fermions contained in $\Theta$
will be
\begin{equation}
  Q_e = \frac{e}{2} \left( -iG_7 G_8 -i \G_5 \G_6 \otimes \one \right) =
 \begin{pmatrix} +e1_4 & & & \cr {}& 0 & & \cr {}&{}& 0&{} \cr
 {}&{}&{}&-e1_4 \end{pmatrix}
\end{equation}
which could be interpreted as follows: if the charged partner of the
baryon doublet $N_1$ (the proton) has the charge $+e$, the charged
partner of the lepton doublet $N_2$ (the electron) should have
the charge $-e$, as it happens, in fact, in nature.

\subsection{THE SYMMETRY $SU(2)\otimes U(1)$} \label{sec82}

 If $\Theta=\begin{pmatrix} N_1 \cr N_2 \end{pmatrix}$ is assumed to
 be in the Pauli representation with $N_1$ and $N_2$ Pauli spinors of
 $\Cl_0(6,1)$ the generators $G_\alpha$ may be assumed to be:
\begin{equation}
\begin{split}
 G_\mu &= \s_3 \otimes 1_2 \otimes \g_\mu \\
 G_{5,6,7} &= \s_2 \otimes \s_{1,2,3} \otimes 1_4 \\
 G_8 &= \s_1 \otimes 1_2 \otimes 1_4 \\
 G_9 &= \s_3 \otimes 1_2 \otimes \g_5
\end{split}
\label{8.9}
\end{equation}

 In this case eq.~\eqref{8.5} takes the form
\begin{equation}
\begin{split}
  \left( p_\mu \,  1 \otimes \g_\mu + p_9 \, 1 \otimes \g_5 + p_{10}
  \right) N_1 + \left( p_8 -i \vec{p} \cdot \vec{\s} \right) N_2 &=0\,, \\
  \left( -p_\mu \,  1 \otimes \g_\mu - p_9 \, 1 \otimes \g_5 + p_{10}
  \right) N_2 + \left( p_8 +i \vec{p} \cdot \vec{\s} \right) N_1 &=0 \,,
\end{split}
\label{8.10}
\end{equation}
where $\vec{p} \cdot \vec{\s}= p_5\s_1 + p_6\s_2+p_7\s_3$. Then
setting
\begin{equation}
 \left( p_8 \pm i \vec{p} \cdot \vec{\s} \right) = \rho e^{\pm
 \frac{i}{2} \vec{\omega} \cdot \vec{\s}} \label{8.11}
\end{equation}
where $\rho =\sqrt{p_5^2+p_6^2+p_7^2+p_8^2}$, eq.~\eqref{8.10} may be
written in the form:
\begin{equation*}
\begin{split}
  \left( p_\mu \, 1 \otimes \g_\mu + p_9 \, 1 \otimes \g_5 + p_{10}
  \right) e^{ \frac{i}{2} \vec{\omega} \cdot \vec{\s}} N_1 + \rho N_2
  &=0\,, \\ \left( -p_\mu \, 1 \otimes \g_\mu - p_9 \, 1 \otimes \g_5 +
  p_{10} \right) N_2 + \rho e^{ \frac{i}{2} \vec{\omega} \cdot \vec{\s}}
  N_1 &=0 \,,
\end{split}
\end{equation*}
in which the $N_1$ doublet manifests an invariance for the phase
transformation:
\begin{equation*}
  N_1 \to e^{ \frac{i}{2} \vec{\omega} \cdot \vec{\s}} N_1
\end{equation*}
which appears manifestly as $SU(2)$ covariance for rotations in
the unit sphere $u_5^2+u_6^2+u_7^2 = 1$, where $\vec{\omega} =
\frac{\vec{p}}{\vert p\vert}$, with $\vert p \vert
=\sqrt{p_5^2+p_6^2+p_7^2}$. Clearly for local transformation this
will give origin to a non abelian Yang-Mills gauge field theory.
$\Theta$ will then present an $SU(2) \otimes U(1)$ internal
symmetry, origin of strong interactions, for the $N_1$ doublet as
seen in eq.~(6.5) which will be absent for $N_2$,
representing the lepton doublet, in which however the geometric
structure of the electroweak model will be present as shown in
section \ref{sec5.2}.

\section{THIRTY TWO COMPONENT SPINORS}

 Continuing the construction, the $P_\alpha$ of eqs.~\eqref{8.3} and
 \eqref{8.4} are a particular realization of the following:
\begin{equation}
 P^\pm_\alpha = \Phi^\dagger \gg_0 \gg_\alpha \left( 1 \pm \gg_{11} \right)
 \Phi \qquad \alpha=1,2,\dots,10\,,
 \label{9.1}
\end{equation}
where  $\Phi$ is a thirty two component Dirac spinor of
end$\Cl(9,1)$ of which $\gg_\alpha$ are the generators and
$\gg_{11}$ the volume element. Then, for $\Phi$ simple (as Weyl
of $\Cl_0(11,1)$)
\begin{equation}
 P_\alpha = P_\alpha^+ + P_\alpha^- = \Phi^\dagger \gg_0 \gg_{\alpha} \Phi
 \label{9.2}
\end{equation}
together with
\begin{equation}
  P_{11} = \Phi^\dagger \gg_0 \gg_{11} \Phi \qquad  P_{12} =
  \Phi^\dagger i\gg_0 \Phi  \label{9.3}
\end{equation}
build up the components of a 12 dimensional real null vector defining
the Cartan's equation
\begin{equation}
  \left( P_A \gg^A + P_9 \gg_9 +  P_{10} \gg_{10} +  P_{11}
  \gg_{11} + i P_{12} \right) \Phi =0 ,
\end{equation}
where $ A=1,2,\dots,8 $.\\ \\
For $\Phi$, in the Dirac spinor representation
\begin{equation}
  \Phi = \begin{pmatrix}  \Theta_1 \cr \Theta_2 \end{pmatrix}
 \label{9.5}
\end{equation}
where $\Theta_1$ and $\Theta_2$ may be considered  as Dirac spinors of
$\Cl(7,1)$, eq.~(9.4) will have the form
\begin{equation}
\begin{split}
  \left( P_\mu G^\mu + P_{11} G_9 + i P_{12} \right) \Theta_1 + G_9\left(
  P_9 - i P_{10} \right) \Theta_2 &=0  \\
  \left( P_\mu G^\mu - P_{11} G_9 + i P_{12} \right) \Theta_2 + G_9\left(
  P_9 + i P_{10} \right) \Theta_1 &=0
\end{split}
\label{9.6}
\end{equation}
and defining
\begin{equation}
  \left( P_9 \pm i P_{10} \right) = \rho e^{\pm i \frac{\sigma}{2}}
  \label{9.7}
\end{equation}
 where $\rho = \sqrt{P_{9}^2 +P_{10}^2}$, eq.~\eqref{9.6} may be cast
 in the form
\begin{equation}
\begin{split}
  \left( P_\mu G^\mu + P_{11} G_9 + i P_{12} \right) e^{i \frac{\s}{2}}
  \Theta_1 + \rho G_9\Theta_2 &= 0 \\ \left( P_\mu G^\mu - P_{11} G_9
  + i P_{12} \right) \Theta_2 + \rho e^{i \frac{\s}{2}} G_9\Theta_1 &= 0
\end{split}
\label{9.6p} \tag{\ref{9.6}$'$}
\end{equation}
which presents an $U(1)$ invariance of $\Theta_1$ generated by
$\gg_9 \gg_{10}$, corresponding to a rotation through an angle
$\s$ in the circle $P_{9}^2+P_{10}^2=\rho^2$. We may interpret it
as the $U(1)$ corresponding to a strong charge (or hyper\-charge).
Then the quadruplet of fermions contained in $\Theta_1$ and
$\Theta_2$ could represent baryons and leptons respectively.

Observe that each one of them obeys in principle to an equation
like \eqref{8.5}. Furthermore, because of the notorious
periodicity theorem of Clifford algebras $\Cl(k,l)$ we have
\begin{equation}
   \Cl( l+4,k+4 ) = \Cl(l+8,k) = \Cl(l,k+8) = \Cl(l,k) \otimes
   {\mathcal R}(16) \label{9.8}
\end{equation}
where ${\mathcal R}\left(16\right)$ stands for the algebra of $16
\times 16$ real matrices. Therefore, since neither $\Cl(8,0)$ nor
$\Cl(0,8)$ may be associated with real spinors, the final Clifford
algebra to study in our construction will be:
\begin{equation}
  \Cl(9,1) = {\mathcal R}(32) =   \Cl(1,9)\,, \label{9.9}
\end{equation}
admitting real Majorana-Weyl spinors, since after this the
geometrical structures, because of the periodicity theorem, will
repeat themselves.

 Before examining the possibility of representing with $\Theta_1$ and
 $\Theta_2$ baryons and leptons we wish first to define dimensional
 reduction procedure in our formulation and to study its meaning.

\section{DIMENSIONAL REDUCTION} \label{sec10}

In our constructive approach, in which at each step  we doubled
the dimension of our spinor space, while we increased by 2 the
dimension of the correlated vector space, dimensional reduction
will simply consist in reversing of those steps.

Precisely, if $\Psi$ is a $2^n$ component Dirac spinor of the
endomorphisms space of a certain Clifford algebra $\Cl(2n-1,1)$
with generators $\g_a$, $a=1,2,\dots,2n$, the dimensional
reduction is operated first in spinor space by the chiral
projectors $\frac{1}{2} \left( 1 \pm \g_{2n+1} \right)$, that is
\begin{equation}
   \Psi \to \vf_{\pm} = \frac{1}{2} \left( 1 \pm \g_{2n+1} \right)
   \Psi \label{10.1}
\end{equation}
where $\vf_\pm$ are the $2^{n-1}$ component Weyl spinors of
$\Cl_0(2n-1,1)$. Correspondingly the $2n+2$ vector space whose
vectors with components $P_A$ are constructed bilinearly from
$\Psi$:
\begin{equation}
  P_A = \Psi^\dagger \g_0 \g_A \Psi\,,  \qquad A=1,2,\dots,2n+2
  \label{10.2}
\end{equation}
with $\gamma_{2n+2}=i\one$, will become, after the spinor-space
dimensional reduction~\eqref{10.1}:
\begin{equation}
  P_A \to p_a^\pm = \Psi^\dagger \g_0 \g_a \left(1 \pm \g_{2n+1}
  \right) \Psi \,, a=1,2,\dots,2n , \label{10.3}
\end{equation}
They span a $2n$ dimensional vector space since, after the spinor
space dimensional reduction, two of the $2n+2$ components $P_A$
and precisely $P_{2n+2}$ and $P_{2n+1}$, will vanish identically,
and dimensional reduction in vector space will be the consequence
of dimensional reduction in spinor space. But then, the spinor so
reduced, will obey an equation in the lower dimensional
momentum-space.

As an example, for $n=2$, consider the case of four component
spinors of Chapter 5. It is easily seen that correspondingly to
the chiral projection:
\begin{equation*}
  \psi \to \frac{1}{2} \left(1 \pm \g_5 \right) \psi = \vf_\pm\,,
\end{equation*}
which reduces the four component spinor $\psi$ to the two
component $\vf_\pm$, the null six-vector $p_a= \left\{
p_\mu,p_5,p_6 \right\}$ studied in Chapter 5, reduces to a null
four-vector
\begin{equation*}
   p_\mu^\pm = \tilde{\psi} \g_\mu \left(1 \pm \g_5 \right) \psi \,,
   \qquad \mu=0,1,2,3\,,
\end{equation*}
since $p_5^\pm =p_6^\pm = \tilde{\psi} \left(1 \pm \g_5 \right)
\psi \equiv 0$.

For $n\geq 3$ the missing terms will generally mean missing of
interaction terms in the equations of motion, and the result of ``dimensional
reduction'' simply consists in ``decoupling'' of the equations of motion.

Obviously depending on the representation of the spinor $\Psi$
and of the $\gamma$-matrices, one may have to use other projectors
in order to reduce by one half the dimension of spinor space. As
an example consider the case of the nucleon doublet dealt with in
section \ref{sec5.1}, where $N=
\begin{pmatrix} \psi_1 \cr \psi_2 \end{pmatrix}$ is a doublet of Dirac
$\Cl(3,1)$ spinors: $\psi_1$ and $\psi_2$. Then, the projectors
read $\frac{1}{2} \left(1 \pm i\G_5 \G_6 \right)$ and it is
easily seen that these projectors send the 8-dimensional null
vector $p_A$. given by eq.~\eqref{5.2} into the six dimensional
one:
\begin{equation*}
  p_a^\pm = N^\dagger \G_0 \G_a \left(1 \pm i\G_5 \G_6 \right) N\,, \qquad
  a=0,1,2,3,7,8 \,.
\end{equation*}
 Since $p_{5,6}^\pm = N^\dagger \G_0 \G_{5,6} \left(1 \pm i\G_5 \G_6
 \right) N \equiv 0$. Also the equation (6.5)
 will correspondingly be reduced, missing the terms $p_5$, $p_6$,
 representing strong interaction terms.

 In general, depending from the representation of the generators
 $\gamma_a$ of $\Cl(2n-1,1)$ one may build up more projectors in
 spinor-space each of which will determine a corresponding
 dimensional reduction in momentum space. Since the $\gamma_a$
 representation determines the representation of the spinor $\Psi$
 above, to each such representation there will correspond a
 dimensional reduction.

 In order to determine them, let us first remember that, for the
 physical interpretation, we need to know the transformation
 properties both of the spinor-multiplet as well of its
 components, with respect to the Poincar\'e group. To this end we
 need then to determine the form of the first four generators
 $\Gamma_\mu$ of the concerned Clifford's algebra; that is
 $\Gamma_1,\Gamma_2,\Gamma_3$ and $\Gamma_0$; assuming they
 represent space- and time-directions, respectively.

 Now the generators $\Gamma_\mu, (\mu =1,2,3,0)$ of $\Cl(2n-1,1)$
 may have, as known, the following four representations:
\begin{equation}
\begin{split}
\text{Weyl:\hspace{3em}}
  \G_\mu^{(1)}  &= \s^1\otimes \gamma_\mu = \begin{pmatrix}
  0& \gamma_\mu\cr \gamma_\mu &0\cr \end{pmatrix}\\
 \G_\mu^{(2)}  &= \s^2\otimes \gamma_\mu = \begin{pmatrix}
  0& -i\gamma_\mu\cr i\gamma_\mu &0\cr \end{pmatrix}\\
\text{Pauli:\hspace{3em}}
 \G_\mu^{(3)}  &= \s^3\otimes \gamma_\mu = \begin{pmatrix}
   \gamma_\mu &0\cr 0&-\gamma_\mu \cr \end{pmatrix}\\
\text{Dirac:\hspace{3em}}
 \G_\mu^{(0)}  &= \one\otimes \gamma_\mu = \begin{pmatrix}
 \gamma_\mu&0\cr 0&\gamma_\mu \cr \end{pmatrix}\\
\end{split}
\end{equation}
where $\gamma_\mu$ are the first four generators of $Cl(2n-3,1)$.

Corresponding the spinor $\Psi$ associated with  $\Cl(2n-1,1)$
may be considered as a doublet of Weyl
(for $\G^{(1)}_\mu ,\g^{(2)}_\mu )$; Pauli (for $\G_\mu^{(3)}$) or
Dirac (for
$\G_\mu^{(0)}$) spinors.

For the Clifford algebra $\Cl(1,2n-1)$ we would have had instead
the generators:
$$\G_\mu^{(j)}=-i\s^j\otimes\gamma_\mu ;\ \
\G_\mu^{(0)}=\one\otimes\g_\mu,\quad j=1,2,3 \eqno(10.4')
$$

Now $-i\s^j$ are the known representation of quaternion imaginary
units. Therefore, the four representation above may be conceived
correlated with quaternion numbers. Now, starting from (10.4) or
(10.4$'$) one may easily complete the representation of the
generators of $\Cl(2n-1,1)$ or $\Cl(1,2n-1)$ and we find that the
corresponding projects are:
\begin{eqnarray}
\text{Weyl:\hspace{3em}}\hfill &&\pi_{1,2}=\frac{1}{2}\left(
1\pm\G_{2n+1}\right),\ \ \text{for}\ \ \G^{(1)}_\mu\ \
\text{and}\ \
\G^{(2)}_\mu ,\nonumber \\
\text{because of which:\hspace{3em}}\hfill
&&P_{2n+2}\equiv 0\equiv P_{2n+1} ;\nonumber\\
\text{Pauli:\hspace{3em}}\hfill
 &&\pi_{3}=\frac{1}{2}\left( 1\pm i\G_{2n}
\G_{2n+1}\right),\ \ \text{for}\ \ \G^{(3)}_\mu ,\nonumber\\
\text{because of which:\hspace{3em}} \hfill
&&P_{2n+1}\equiv 0\equiv P_{2n} ;\\
\text{Dirac:\hspace{3em}}\hfill
&&\pi_{0}=\frac{1}{2}\left( 1\pm
i\G_{2n-1}
\G_{2n}\right),\ \ \text{for}\ \ \G^{(0)}_\mu ,\nonumber \\
\text{because of which:\hspace{3em}} \hfill
&& P_{2n}\equiv 0\equiv P_{2n-1}\nonumber
\end{eqnarray}
The Dirac case may be further extended if one supposes that in
$\G^{(0)}_\mu$ of eq. (10.4) also the four $\gamma_\mu$ are in the Dirac
representation $\gamma^{(0)}_\mu$; we will indicate it with:
\begin{equation}
\G^{(00)}_\mu = \one\otimes\gamma^{(0)}_\mu = \begin{pmatrix}
\g^{(0)}_\mu &0\cr 0&\g^{(0)}_\mu\cr \end{pmatrix}
\end{equation}
and the corresponding projector will be:
\begin{equation}
\pi_{00} =\frac{1}{2}\left(1\pm i\G_{2n-3}\G_{2n-2}\right) ,
\end{equation}
because of which
\begin{equation}
P_{2n-2}\equiv 0\equiv P_{2n-3}
\end{equation}

This may be continued up to $\G^{(0\dots 0)}_\mu$ with $m$ zeros,
which we will indicate with $\G^{(0m)}_\mu$ and which will have
the corresponding projector:
\begin{equation}
\pi_{0m} =\frac{1}{2}\left(1\pm i\G_{2n-2m+1}\G_{2n-2m+2}\right)
\end{equation}
because of which
\begin{equation}
P_{2n-2m+2}\equiv 0\equiv P_{2n-2n+1}
\end{equation}

It is easily seen that, given $n$, the maximal $m$ to be considered
is $m=n-2$ by which the original $2^n$ Dirac spinor splits in
$2^m$ Dirac spinors each one with 4 components.

 When working in space-time in the traditional method, the extra
 dimensions are introduced in order to explain multiplicities and
 internal symmetry of fermions in their equations of motions or Lagrangians,
 and are afterwards eliminated by dimensional reduction, confining
 them in compact manifold of unobservable size.

 Here, instead, the extra dimensions appear as
 extra terms in the equations of motion in momentum space and, for $n>2$, they
 acquire then the meaning of terms representing the interaction of
 fermions with external fields, as in eq.~(6.5). Their
 elimination may, then, be interpreted as a decoupling, or as
the reduction of the
 equations of the concerned fermions in absence of those interaction
 terms, a quite natural interpretation since it is natural that a
 proton (represented by $\psi_1$ in (6.5) say), being far from a neutron
 (represented by $\psi_2$, say), like those of cosmic rays, when traveling
 in empty space say, will
 simply obey the Dirac equation, for a charged fermion.

 \section{BARYONS AND LEPTONS} \label{sec11}

 \subsection{THE BARYON MULTIPLET}

 Let us now assume in eq. (9.5) the Dirac basis, the 16 component spinor
 $\Theta_1$ to be of the form:
\begin{equation}
 \Theta_1 = \begin{pmatrix} q_1 \cr q_2 \cr q_3 \cr q_4 \end{pmatrix}: =
 \Theta_B \label{11.1}
\end{equation}
where $q_j$ are Dirac spinors, representing a quadruplets of
fermions or quarks, presenting strong charge represented by its
$U(1)$ covariance as seen in eq.~(9.6$'$). It obeys
eq.~\eqref{8.5} and $P_\alpha$ given by \eqref{8.3}, \eqref{8.4},
being the components of a null vector satisfy identically to (we adopt
$\Cl(1,9))$:
\begin{equation}
 P_\mu P^\mu = P_5^2+P_6^2+P_7^2+P_8^2+P_9^2+P_{10}^2=M^2 ,
 \label{11.2}
\end{equation}
whose directions define $S^5$. Therefore it is to expected that the
quadruplets may present a maximal $SU(4)$ internal symmetry
(covering group of $SO(6)$), which will be the obvious candidate
for flavor internal symmetry. The most straightforward way to set
it in evidence is to determine the 15 generators if the Lie algebra
of $SU(4)$ (determined by the commutators $\left[G_i,G_k\right]$
for $5 \leqslant j,k \leqslant 10$) represented by $4 \times 4$
matrices, whose elements are either $1_4$ or $\g_5$, acting in
the space of the $\Theta_B$ quadruplet. Let them be $\lambda_j$; where
$1\leqslant j \leqslant 15$. Denote with $f_j$, $1\leqslant j
\leqslant 15$ the components of a tensor building up an
automorphism space of $SO(6)$. Then, a natural\footnote{ In a
similar way as $\left( p_\mu \g^\mu + \left[ \g_\mu ,
\g_\nu\right]F^{\mu\nu} +m \right) \psi =0$, where $F_{\mu\nu}$
is the electromagnetic tensor, is a natural equation for
a fermion, the neutron say, presenting an anomalous magnetic
moment.} equation of motion for $\Theta_B$ could be:
\begin{equation}
  \left( P_\mu G^\mu + \sum^3_{j=1}\lambda_jf^j+\sum_{j=4}^8
\lambda_j f^j + \sum_{j=9}^{15}
  \lambda_j f^j \right)\Theta_B =0
\end{equation}
 where $\lambda_1\lambda_2\lambda_3$ are the generators of
$SU(2)$; $\lambda_1,\dots ,\lambda_8$ those of $SU(3)$
and $\lambda_1,\dots ,\lambda_{15}$ those of $SU(4)$.

 In order to study the possible physical information contained in the
 quadruplet $\Theta_B$, let us now operate with our dimensional
 reduction.

 The most obvious will be to adopt the projector $\frac{1}{2} \left(1
 \pm G_9 \right)$ which is the image in $\Cl(9,1)$ of $\frac{1}{2}
 \left(1 \pm \gg_9 \gg_{10}\right)$ of $\Cl(1,11)$. Now, the reduction
\begin{equation}
  \Theta_B \to \frac{1}{2} \left( 1 \pm G_9 \right) \Theta_B = N_\pm
  \,.  \label{11.3}
\end{equation}
implies that
\begin{equation}
  p_{9,10}^\pm = \Theta_B^\dagger G_0  G_{9,10} \left(1 \pm G_9\right)
  \Theta_B \equiv 0 \label{11.3p} \tag{\ref{11.3}$'$} \,.
\end{equation}
 Therefore the null vector $P_\alpha$ given by \eqref{8.3} and \eqref{8.4}
 reduces to $p_A$ given by eq.~\eqref{5.2}, which means that $N_+$ or
 $N_-$, conceived as a doublet of fermions obeys eq.~(6.5) of the
 nucleon doublet.  It
 may be shown that if we substitute in that equation the Dirac spinors
 $\psi_1$,$\psi_2$ with Pauli spinors and we operate with the projector $\pi_3$
 of Chapter 10 in signature
 $(1,7)$, the intermediate terms acquires a factor
 $i 1_4$, instead of $\g_5$, that is eq.~(6.5$'$) becomes a
 quaternionic equation of the form:
\begin{equation}
  \left( p_\mu \cdot 1 \otimes \gamma^\mu + i \vec{\pi} \cdot
  \vec{\s}+
  M \right) N =0 \label{11.5}
\end{equation}

 It is easy to see, that if we operate the dimensional reduction with the
 projectors $\frac{1}{2} \left(1 + G_7 G_8 \right)$, then $p_{7,8}
 \equiv 0$,  and again the remaining equation
 is a quaternionic one.

If we now compare these results with the general eq.~(11.3) we see that,
in all these cases of dimensional reduction, only the internal symmetry
represented by the first sum containing $\lambda_1,\lambda_2,\lambda_3$
generators of $SU(2)$ isospin has emerged; which had to be expected in fact,
since in the construction of our higher dimensional spinor spaces and of the
corresponding vector spaces, where to interpret Cartan's equations as equations
of motion in momenta space, the signature resulted unambiguously defined to
remain Lorenzian, which in turn determined the emergence of the fundamental
role of the quaternion field of numbers. These in turn are at the origin of
$SU(2)$ isospin symmetry of the strong nuclear forces, which in fact rules all
low energy dynamical phenomena in nuclear physics. This however could not be
the end of the story, since, dealing with $\Cl(1,9)$ notoriously associated
with octonions, presenting the automorphism group $G_2$, one could expect to
find equations presenting an
$SU(3)$ group of symmetry, subgroup to $G_2$, represented in eq.~(11.3) by the
first two sums, as in fact observed in higher energy strong interaction
phenomena, when new terms of interaction appear in the equations of motion as
will be shown in Chapter 12.

\subsection{THE LEPTON MULTIPLET} \label{sec12}

 The multiplet $\Theta_2$ should then not possess strong charge
 generated by $\gg_9\gg_{10}$; we will then interpret it as
 representing leptons:
\begin{equation}
  \Theta_2 = \begin{pmatrix} \ell_1 \cr \ell_2 \cr \ell_3 \cr \ell_4
  \end{pmatrix} := \Theta_L \label{12.1} \,.
\end{equation}
Since now the strong charge is missing, $\Theta_L$ will not obey
to eq. (8.5) which implies the possibility of eq.(11.3) and the
presence of strong interactions.

The absence of strong charge will impose at least one step of our
dimensional reduction: two terms will be missing from eq. (8.5)
valid for $\Theta_B$, while the spinor $\Theta_L$ will reduce to
an 8-component spinor which will be indicated with $L$. In order to perform this dimensional
reduction we will adopt the rules of Chapter 10 for $n=4$, since
$\Theta_L$ may be associated with $\Cl(1,7)$. Then we may think
their generators to have the form (10.4$'$) where $\g_\mu$ are the
standard Dirac $4\times 4$ matrices and the projectors are given
by (10.5) with $n=4$. They are:
\begin{eqnarray*}
\text{Weyl:\hspace{3em}}\hfill &&\pi_{1,2}=\frac{1}{2}\left(
1\pm G_{9}\right)\\
\text{because of which:\hspace{3em}} &&\hfill P_{10}\equiv 0\equiv P_{9} \\
\text{Pauli:\hspace{3em}}\hfill &&\pi_{3}=\frac{1}{2}\left( 1\pm iG_{8}
G_{9}\right)\\
\text{because of which:\hspace{3em}} \hfill &&P_{9}\equiv 0\equiv P_{8} \\
\text{Dirac:\hspace{3em}} \hfill &&\pi_{0}=\frac{1}{2}\left( 1\pm G_{7}
G_{8}\right)\\
\text{because of which:\hspace{3em}} \hfill &&P_{8}\equiv 0\equiv P_{7}\\
\end{eqnarray*}

The projectors $\pi_{1,2,3}$ and $\pi_0$ will reduce $\Theta_L$ to
the 8-component spinors $L^{1,2,3}$ and $L^0$ which will be
Weyl-, Pauli- and Dirac-spinors, respectively and the $\G_\mu$
matrices will be the ones of (10.4$'$). For each of them the
momentum space will be 8-dimensional, reduced from the original
ten-dimensional one satifying eq.(11.2) which for $L^{1,2}$
will reduce to:
\begin{equation*}
p_\mu p^\mu = p^2_5+p_6^2 + p_7^2+p_8^2 = m_{1,2}^2
\end{equation*}
for $L^3$ will reduce to:
\begin{equation}
  p_\mu p^\mu = p^2_5+p_6^2 + p_7^2+p_{10}^2 = m_3^2
  \end{equation}
and for $L^0$ will reduce to:
\begin{equation*}
p_\mu p^\mu = p^2_5+p_6^2 + p_9^2+p_{10}^2 = m_0^2\,.
\end{equation*}
Which means, comparing with (11.2), that in principle the masses
of the leptons will be lower than those of the baryons, and
different among themselves.

Observe further that from what was discussed in section 6.3, in
each doublet one lepton should be electrically charged and one
neutral.

The lepton multiplet $\Theta_L$ should then reduce to 3 doublets
$L^1,L^2,L^3$, like the $N_2$-doublets defined in Section 8.1, labeled by the
quaternion imaginary units
plus $L^0$. If $L^1$ and $L^2$ doublets are considered
equivalent then we would have only 3 non-equivalent pairs of
charged-neutral leptons which might be correlated with the three
families of leptons which have been observed in nature.

The one step of dimensional reduction, imposed by the absence of strong
charge for the lepton quadruplet $\Theta_L$, will in any case generate the
emergence of an $SU(2)$ internal symmetry, which for leptons could be
envisaged in the one of the electroweak model discussed in section 6.2.

Up to now we operated with dimensional reduction on the quadruplet $\Theta_B$
and $\Theta_L$ separately. However it is more natural to operate the reduction
directly on the 32-component spinors studied in Chapter 9 and we will obtain
the 16-component spinors of Chapter 8.

\subsection{THE BARYON-LEPTON QUADRUPLETS}

Let us suppose that in the spinor $\Phi =\begin{vmatrix}
\Theta_1\\ \Theta_2\end{vmatrix}$ of the antomorphism space of
$\Cl(1,11)$ in eq. (9.5), $\Theta_1$ and $\Theta_2$ which we
identified as the baryon and lepton multiplets $\Theta_B$ and
$\Theta_L$ respectively, are of the form
\begin{equation}
\Theta_B = \begin{vmatrix} \Theta_{b_1}\\
\Theta_{b_2}\end{vmatrix} ,\quad
\Theta_L = \begin{vmatrix} \Theta_{\ell_1}\\
\Theta_{\ell_2}\end{vmatrix}
\end{equation}
with $\Theta_{b_j}$ and $\Theta_{\ell_j}$ Dirac spinors of
$\Cl(1,7)$. Then, remembering that dimensional reduction of the lepton
quadruplet $\Theta_L$  is necessary because of the absence of strong charge, we
could reduce the spinors space with the
projector
\begin{equation}
\pi_{00} =\frac{1}{2}\left(1+\gg_7\gg_8\right)
\end{equation}
by which $P_7\equiv 0\equiv P_8$ and the spinor $\Phi$ would
reduce to
\begin{equation}
\Phi = \begin{vmatrix} \Theta_B\\
\Theta_L\end{vmatrix} \longrightarrow\begin{vmatrix}
\Theta_{b_1}\\ 0\\ \Theta_{\ell_1}\\ 0\end{vmatrix} =\Theta
\end{equation}
that is to a sixteen component baryon-lepton doublet like those
considered in section 8.1:
\begin{equation*}
\Theta = \begin{vmatrix} N_1\\ N_2 \end{vmatrix} = \begin{vmatrix}
\psi_1\\ \psi_2 \\ \psi_3 \\ \psi_4 \end{vmatrix}
\end{equation*}
in which $\psi_1,\psi_2$ are baryons: we will indicate them with
$b_1,b_2$ and $\psi_3,\psi_4$ leptons to indicate with
$\ell_1,\ell_2$. We have seen in the preceding section 11.2 that
we may obtain them after dimensional reduction from $\Theta_B$
and $\Theta_L$ in $3+1$ ways through the projectors
$\pi_{1,2,3,0}$ labeled according to the directions of
quaternion field of numbers. Therefore the resulting baryon-lepton quadruplet
will
have the form:
\begin{equation}
\Theta^{(j)} =\begin{pmatrix} b^{(j)}_1\\ b^{(j)}_1\\ \ell^{(j)}_2\\
\ell^{(j)}_2 \end{pmatrix} ,\quad j=1,2,3,0
\end{equation}

The form (11.11) is suggestive since it is apt to set in evidence
several of the common properties of baryons and lepton in
particular the natural arising of $SU(2)\otimes U(1)$ internal
symmetry.

We have to remind that while leptons $\ell_1,\ell_2$ are not
subject to strong interactions and therefore for them the
dimensional reduction from $\Theta_L$ to $\Theta_{\ell_1}$ (or
$\Theta_{\ell_2}$) is necessary, for the baryons $b_1,b_2$ of the
doublet, it is not so since they may be subject to strong
interactions. In particular it is known that these interactions
present an $SU(3)$ symmetry, and the $SU(3)$ group is notoriously
a subgroup of $G_2$, the automorphism group of octonions.

We have seen up to now that quaternions seem to play a basic role
in the interpretation of some elementary physical phenomena
concerning fermions, which may be derived from the geometry of
simple or pure spinors; we may then reasonably expect that also
octonions or Cayley numbers might play a role, specially because
we ended up to deal with spinors associated with the algebras:
$\Cl(1,1+8)=\Cl(1+8,1)$, notoriously associated with octonions.

\section{THE OCTONION FORMALISM}\label{sec13}

 Suppose now the 16 component spinor $\Theta$  to be a doublet of
 $\Cl_0(5,1)$  Weyl spinors $\theta_1$,$\theta_2$: $\Theta = \begin{pmatrix}
\theta_1 \cr \theta_2
 \end{pmatrix}$. Then $G_A$, generators of $\Cl(1,7)$, may have the form
\begin{equation}
  G_a = \begin{pmatrix}
  0 & -i \G_a \cr i \G_a & 0 \end{pmatrix} ;
  \ \ G_8 = \begin{pmatrix} 0 & 1 \\ 1 & 0 \end{pmatrix} ;
    G_9 = \begin{pmatrix} 1 & 0 \\ 0 & -1 \end{pmatrix}
  \ \ a= 1,2,\dots,7 \,.
 \label{13.1}
\end{equation}
 and eq.~\eqref{8.5}, taking into account of eq.~\eqref{9.6},
reads\footnote{ Eq.~\eqref{13.2} may be easily formulated for the
signature (1,9); one needs only to substitute $\pm iP_{10}$,$P_9$
and $i \G_a$ with $\pm P_{10}$, $iP_9$ and $\G_a$ respectively,
$\G_a$ being the generators of $\Cl(1,5)$.}:
\begin{equation}
  P \Theta = \begin{pmatrix} \pm iP_{10} + P_9 & P_8 -i \sum_{j=1}^7 P_j
  \G_j \cr P_8 + i \sum_{j=1}^7 P_j \G_j & \pm i P_{10} -P_9 \end{pmatrix}
  \begin{pmatrix} \theta_1 \cr \theta_2 \end{pmatrix} =0 \,.
  \label{13.2}
\end{equation}
A number of authors~\cite{19} have proposed to adopt the formalism of
octonion division algebra for the study of physics in ten dimensional
momentum spaces of signature $(1,9)$ and have adopted the equation:
\begin{equation}
P\Theta = \begin{pmatrix} p_{10}+p_9 & p_8 - \sum_{j=1}^7 p_j e_j \cr
                  p_8 + \sum_{j=1}^7 p_j e_j & p_{10}-p_9
  \end{pmatrix}
  \begin{pmatrix}
    0_1 \cr 0_2
  \end{pmatrix} =0\,,
  \label{13.3}
\end{equation}
where $e_1,e_2,\dots,e_7$ are the anticommutative imaginary units
of octonions and $\Theta=\begin{pmatrix}0_1 \cr 0_2
\end{pmatrix}$ is a two component spinors with $0_1$ and $0_2$
representating octonions.

Both eqs. (12.2) and (12.3) contemplate the same signature ((9,1)
or (1,9)) however while eq. (12.2) implies:
\begin{equation}
   P_0^2 -P_1^2 -\dots - P_9^2 -P_{10}^2 =0 \,, \label{13.2p}
   \tag{\ref{13.2}$'$}
\end{equation}
eq. (12.3) implies:
\begin{equation}
   p_{10}^2-p_9^2 -\dots - p_1^2-p_0^2  =0 \,, \label{13.3p}
   \tag{\ref{13.3}$'$}
\end{equation}
that is the time components $p_0$ and $p_{10}$ are interchanged
(apart from the $\pm$ sign in front of $P_{10}$ deriving from eq.
(9.4) and (9.6)). This means that if in eq. (12.2) we perform the Wick
rotations:
\begin{equation*}
  i P_{10} \to P_{10} \,; \qquad P_0 \to i P_0
\end{equation*}
eq. (12.2) and eq. (12.3) must be equivalent. In fact it is
known~\cite{20} that ${\mathop {\rm Spin}} (9,1)={\mathop {\rm
Spin}} (9,1)={\mathop {\rm SL}}(2,0)={\mathop {\rm SL}}(32, \R)$
where $0$ stands for octonions.

Recently T. Dray and C.A. Manogue \cite{19}, starting from eq.
(12.3) have proposed a dimensional reduction from 10 to 4
dimension in momentum space, adopting the map
$$
\pi (q) =\frac{1}{2} (q+\ell q\ell )
$$
where $q$ represents an octonion and $\ell$ a preferred octonion
imaginary unit (they use $\ell =e_7$) In this way they obtain
from eq. (12.3), spinor equations in 4-dimensional momentum space
apt to represent 3 generations of lepton pairs -- one massive and
one massless in each pair -- each generation corresponding to one
of the 3 quaternion imaginary units.

Compared to the octonions eq. (12.3) the advantage of eq. (12.2)
is that it is straightforward to
 read the physical meaning of the spinor $\Theta$, especially when we
 perform dimensional reduction step by step, and furthermore the
 algebra of the non diagonal terms in eq. (12.2) is the familiar one of Clifford
 algebras, while that in eq. (12.3) is the non-associative one of
 octonions. However, in a similar way as the algebra of quaternions
 seems to play an important role in the explanation of the geometrical
 structure of some elementary physical phenomena, in particular they
 might be at the origin of $SU(2)\otimes U(1)$ internal symmetry, and of
 the 3 families of baryon-lepton quadruplets,
 the formulation
 of eq. (12.3) in terms of octonions could
 be helpful for understanding further aspects of the geometrical
 structure of some elementary physical phenomena, like the ${\mathop
 {\rm SU}}(3)$ (subgroup of $G_2$) internal symmetry.

To set this
 in
 evidence we should then try to correlate eqs. (12.2) and
 (12.3), that is, to express octonions in terms of Clifford
 algebra elements. This seems to be indeed possible in the frame of
 $\Cl(2,3)$ and will be extensively analyzed elsewhere \cite{21}.
 For the moment we will tentatively mention some preliminary
 results which are coherent with our previous study.

Let us take for the generators of $\Cl(2,3)$ the $4 \times 4$
 matrices
\begin{equation}
  \gamma_n = \begin{pmatrix} 0 & -\s_n \cr \s_n & 0
  \end{pmatrix} \,; \quad \gamma_0 = \begin{pmatrix} 0 & 1 \cr
  1 & 0 \end{pmatrix} \,; \quad \gamma_5 = \begin{pmatrix} 1
  & 0 \cr 0 & -1 \end{pmatrix} \,; \quad n=1,2,3 \label{13.4}
\end{equation}
then we may define the octonion units with
\begin{equation}
   e_n=\gamma_n \,;\quad e_7= i \gamma_5 \,; \quad e_{n+3} = e_n e_7 =
   i \gamma_n \gamma_5\,; \qquad n=1,2,3\,.
\end{equation}
which satisfies the multiplication rules of octonions, that is
\begin{multline}
 e_\ell e_m=-\delta_{lm}+
 \varepsilon_{lmn} e_n;\hspace{0.75cm} e_\ell e_{m+3}= - \delta_{lm}e_7 -
 \varepsilon_{lmn} e_{n+3};\\ e_{l+3} e_{m+3} = - \delta_{lm} -
 \varepsilon_{lmn} e_n;
 \end{multline}
where $\varepsilon_{\ell mn}$ is the
emisymmetric tensor,
provided the rule for the multiplication of matrices is
appropriately modified, as proposed by J. Daboul and R. Delbourgo
\cite{22}. It will be shown in ref.[21] that the same result may
also be obtained in the frame of the Clifford algebra $\Cl(2,3)$
maintaining the standard rule for matrix multiplication.

We will show now that, in the frame of $\Cl(1,9)$, there are
several sub-algebras apt to represent octonion units as well as
their multiplication rules after adoption of the mentioned
conventions, and to give origin to equations of motion of the type (11.3) with interaction terms represented by the first two sums.

Let us start from the representation of $G_A$, generators of
$\Cl(1,7)$, as given in eqs. (12.1); inserting in them the
representations of $\G_a$ given in eqs. (6.3) one easily finds:
\begin{equation}
G_{4+n} = \begin{pmatrix} 0 & -\s_n \\ \s_n &0 \end{pmatrix}
\otimes \g_5 :=e_n\otimes\g_5 ;\ \  n=1,2,3
\end{equation}
where we have adopted the representation (12.4), (12.5) of the
octonion units $e_n$. If we further define
\begin{equation}
iG_9 := e_7\otimes 1
\end{equation}
and
$$
iG_{4+n} G_9 := ie_ne_7\otimes \g_5=e_{n+3} \otimes \g_5
\eqno(12.8')
$$
we obtain all the seven imaginary octonion units expressed in the
frame of $\Cl(1,7)$ which, with the mentioned rules, close the
octonion algebra.

Let us now define the complex octonions with:
\begin{equation}
\begin{split}
  \u_0 = \frac{1}{2} \left( 1 - i e_7 \right);\quad
  \u_n = \frac{1}{2} \left( e_n - i e_{n+3} \right);\\
  \Bar{\u}_0= \frac{1}{2} \left( 1 + i e_7 \right);\quad
  \Bar{\u}_n =\frac{1}{2} \left( e_n + i e_{n+3} \right)
\end{split}
 \label{12.9}
\end{equation}
They satisfy:
\begin{equation}
\begin{split}
 \u_0^2 = \u_0 \,; \qquad \u_0 \Bar{\u}_0 =0\,; \qquad\u_n \u_0 =
 \Bar{\u}_0\u_n = \u_n \,; \\ \u_0 \u_n = \u_n \Bar{\u}_0 =0\,; \qquad
 \u_l \Bar{\u}_m = - \delta_{lm}\,; \quad \u_l \u_m =
 \varepsilon_{lmn} \Bar{\u}_n
\end{split}
\label{12.10}
\end{equation}
together with the complex conjugate equations.

 This algebra is known to be invariant under the group $SU(3)$~\cite{23}.
 Precisely $\left( \u_1, \u_2, \u_3 \right)$ and $\left( \Bar{\u}_1,
 \Bar{\u}_2, \Bar{\u}_3 \right)$ transform like $({\bf 3})$ and
 $(\Bar{\bf{3}})$ representations of $SU(3)$, respectively; while
 $\u_0$ and $\Bar{\u}_0$ like singlets. Therefore it manifests the
 $SU(3)$ automorphism of the octonion algebra.

 It is now easy to express, through eqs. (12.7), (12.8) and
 (12.8$'$) the complex octonions $\u_0,\u_n; \Bar{\u}_0,
 \Bar{\u}_n$ in terms of the generators $G_{4+n}, G_9$. With them
 it should then be possible to build up terms transforming as $({\bf 3})$ and
 $(\Bar{\bf{3}})$  of $SU(3)$ and then presenting on $SU(3)$ internal
 symmetry. Observe that these terms contain all the projectors $\frac{1}{2} (1\mp ie_7)=
 \frac{1}{2} (1\pm G_9)$. Now if we act with these projectors in
 spinor space we operate a dimensional reduction, as discussed in
 section 11.1, and the corresponding equation of motion reduces to
 the one of the nuclear doublet presenting on $SU(2)$ isospin
 symmetry.

 Let us now extend the above construction to our Clifford algebra
 $\Cl(1,9)$ by defining its generators $\gg_\alpha$ as follows:
 \begin{multline}
 \gg_\alpha : \gg_A = \begin{pmatrix} 0&-iG_A\\ iG_A &0
 \end{pmatrix} ; \ \
 \gg_9 = \begin{pmatrix} 0&-G_9\\ G_9 &0
 \end{pmatrix}, \ \
 \gg_{10} = \begin{pmatrix} 0&-i\\ i &0
 \end{pmatrix}\\
\gg_{11} =\begin{pmatrix} 1&0\\ 0 &-1
 \end{pmatrix} \qquad A=1,2,\dots ,8
 \end{multline}

 Then, adopting the above representations for the generators
 $G_A$, we easily arrive at the following definitions:
 \begin{equation}
 \begin{split}
 \gg_{6+n} = \begin{pmatrix} 0&-\s_n\\ \s_n &0 \end{pmatrix}
 \otimes \G_7 &:= e_n\otimes \G_7\ \ \ n=1,2,3\\
 i\gg_{11} &:= e_7 \otimes 1\\
 i\gg_{6+n} \gg_{11} &= ie_ne_7 \otimes \G_7 = e_{n+3} \G_7
 \end{split}
 \end{equation}
 which, again, close the octonion algebra with the adoption of
 the mentioned rules.

 In this case the projectors presented by complex octonions will
 be $\frac{1}{2} (1\pm\gg_{11})$ which, in our model of Chapter 9,
 splits the 32-component spinor $\Phi$ in the baryonic quadruplet
 $\Theta_B$ and the leptonic one $\Theta_L$. Therefore the terms
 $\gg_{6+n}(1+\gg_{11}$) may give rise to $SU(3)$ internal symmetry
 for $\Theta_B$ presenting strong charge while it should be absent
 for $\Theta_L$  (where however it could determine the structure
 of the electroweak model determining the value 0.25 of
 $\sin\theta$, see section 6.2)). This possible $SU(3)$ symmetry
 for the baryons contains, as seen above, the isospin $SU(2)$ as a
 subgroup, therefore it would be natural to interpret it as
 flavour internal symmetry, acting on the baryons interpreted as operators,
with  all that which follows.

 Other possible representations of the octonion units may be found
 in the frame of $\Cl(1,9)$. In fact in Chapter 10 we have seen
 that in the frame of $\Cl(1,5)$ the first four generators of the
 algebra may have the form:
 \begin{equation}
 \G^{(n)}_\mu = \s_n\otimes\g_\mu\qquad \mu = 0,1,2,3
 \end{equation}
 with $\g_\mu$ generators of $\Cl(1,3)$, where $n=1,2$
 characterize the spinor associated with $\Cl(1,5)$ as Weyl
 doublets, while $n=3$ as a Pauli doublet. In section 11.3 we have
 seen a natural reduction of the spinor $\Phi$ in 3 families of
 baryon-lepton quadruplets represented by eq. (11.11). For these
 the corresponding first four generators of $\Cl(1,7)$ could have
 the form:
  \begin{equation}
 \G^{(n)}_\mu = \begin{pmatrix} 0&-\s_n\\ \s_n&0 \end{pmatrix}
 \otimes\g_\mu := e_n\otimes\g_\mu
 \end{equation}
 Then with the help of definition (12.8) and of:
$$
i G^{(n)}_\mu G_9 := ie_ne_7\otimes\g_\mu =e_{n+3}\otimes\g_\mu
 \eqno(12.14')
 $$
 we arrive to another representation of the octonion units, out of
 which a complex octonions algebra invariant under the $SU(3)$
 group may be defined (the same may be repeated with the
 generators $\gg_\alpha$).

 A further representation of octonion units and of the
 corresponding complex octonion algebra may be found if we take
 for the generators $\g_\mu$ of $\Cl(1,3)$ the representation:
  \begin{equation}
 \g_\mu : \g_\mu = \begin{pmatrix} 0&-\s_n\\ \s_n&0 \end{pmatrix}
 ,\ \
\g_0 =\begin{pmatrix} 0&1\\ 1&0 \end{pmatrix} ,\ \ n=1,2,3
 \end{equation}

In this case we easily arrive at
\begin{equation}
G^{(j)}_k := e_j\otimes e_k\qquad j,k =1,2,\dots ,6
\end{equation}
and
\begin{equation}
-G^{(7)}_7 := -G_9\otimes \g_5 := e_7\otimes e_7
\end{equation}
which generates, in the complex version, an algebra invariant
under the $SU(3)\otimes SU(3)$, which under certain conditions,
could be candidate for internal symmetry groups. If they could represent also
the color $SU(3)$ as advocated by F. G\"ursey [19], is for the moment an open
question.

In this way the two first sums appearing in the general eq. (11.3) could be
realized. For the possible realization of the last term implying the, in
principle possible, $SU(4)$ internal symmetry in very high energy elementary
particle phenomena, there is no sign for the moment. If the parallelism between
baryons and leptons, as mentioned in section 11.3, is a dominant feature of
elementary particle phenomena then one could expect the discovery of a fourth
neutrino, which in fact is being searched for in some laboratories (Fermilab).

Observe that $SU(3)$ is a subgroup of the automorphism group $G_2$ of octonions
and it emerges when among the octonion units a preferred one is chosen; in our
case $e_7$. This in turn, in our Clifford algebra approach, amounts in choosing
a preferred direction in momentum space corresponding to a preferred generator
of the
concerned Clifford algebra. In the example given
above they have been in turn: $G_9,\gg_{11},\g_5$.
Obviously, also $\G_7$ could have been used. These were the
volume elements of our even dimensional Clifford algebras and, in
the complex octonion algebras, they appear in the form of
generalized chiral projectors like $\frac{1}{2} (1\pm G_9),
\frac{1}{2} (1\pm \gg_{11}), \frac{1}{2} (1\pm \g_5)$. This means
that in the corresponding $SU(3)$ covariant terms in the
Lagrangian densities the corresponding spinors will appear as
chirally projected, which might be at the origin of their
inobservability.

This possibility is not new in physics where, as an example, the weak
current responsible for the neutron decay is $\tilde{P}\g_\mu
(1+\g_5)N$, and we know that the massive left-handed neutron
$(1+\g_5)N$, which appears in the weak interaction term, is not
observable as a free particle since it is not invariant under the
reflections of the Lorentz group (it obeys to coupled Dirac
equations); it behaves as a free particle, only asymptotically in
the limit of high momenta, when its mass may be, with good
approximation, set to zero.

If the unobservability of colored quarks could have a similar origin,
the necessity of conceiving observable fermions as bound states
of quarks could be revised (in the above example the neutron does
not necessitate to be considered as a bound state of its left-
and right-handed Weyl components, which are unobservable as free
fermions).

What emerges from this preliminary study is that the geometrical
structure of the Clifford algebra $\Cl(1,9)$ and of its
endomorphism spinor space is very rich and may naturally
accommodate several features of the observed elementary natural
phenomena of the known fermions.

In particular, all 3 stages of complexity seem to play a
fundamental role: complex numbers as origin of the $U(1)$ group
of charge in the charged-neutral fermion doublets; quaternions as
origin of isospin group of symmetry $SU(2)$ as well as of the 3
families of lepton-baryon doublets, correlated with the 3 units
of quaternions; octonions  as the origin of $SU(3)$ internal
symmetry perhaps including color, again the number 3 being
correlated with the number of quaternion units and also the
number 8 of the eightfold representations of $SU(3)$ with the
periodicity theorem of Clifford algebras and with the fundamental
role of the real algebras $\Cl(1,9)=\Cl(9,1) =R(32)$
representable through octonions in their automorphism space of
Majorana-Weyl spinors.

There are several more geometrical aspects $\Cl(1,9)$ and of its
endomorphism spinor space which should be further analysed
because of their possible correlation with physics. We will try now only
to mention some of them.

\section{FURTHER GEOMETRICAL ASPECTS}

\subsection{SIMPLICITY CONSTRAINTS}

We based our constructive approach, consisting in embedding
spinor spaces and null-vector spaces in higher dimensional ones,
on propositions 1 and 2. Both of them deal with simple (or pure)
spinors as defined by \'E. Cartan \cite{1}. Now, for even
dimensional Clifford algebras $\Cl(2n)$, all Weyl spinors are
simple for $n\leq 3$; while for $n\geq 4$ they are subject to
constraint equations which may be derived from eqs. (2.12).

To start with, for $n=4$, the 8-dimensional Weyl spinors
$\theta_\pm =\frac{1}{2} (1\pm G_9)\Theta$ associated with
$\Cl_0(1,7)$ are simple iff in eq.(2.12) $F_0=0$:
\begin{equation}
F_0 :=\langle\theta^t_\pm \G_0\theta_\pm\rangle = 0
\end{equation}

We know that $\theta_\pm$ may also be conceived as Pauli spinors
of $\Cl(1,6)$ or as Dirac spinors of $\Cl(1,5)$. In this latter
case simplicity eq. (13.1) would imply $M=0$ in eq. (6.5). This condition was
sometimes imposed ad hoc in nucleon-pion physics; here we see that it may be
obtained imposing the semplicity constraint to the nucleon doublet.

The eight dimensional Weyl spinors $\theta_+\theta_-$ of $\Cl(8)$
are specially important since they present a particular auter
automorphism, first discovered by Sturdy \cite{23} and named by
\'E. Cartan \cite{1} the triality principle. In fact the 3 eight
dimensional quadrics
\begin{equation}
Q_1 = \langle B\theta_+,\theta_+\rangle ,\ \ Q_2 = \langle
B\theta_-,\theta_-\rangle ;\ \ Q_3=X_AX^A ,
\end{equation}
where $X_A$ are the orthonormal components of a vector of $\C^8$,
are obviously invariant with respect to the group $O(8)$ and to its
covering group pin (8). The same is obviously true for
\begin{equation}
F=\langle B\theta_+, G_A\theta_-\rangle X^A
\end{equation}

Furthermore, under the transformations:
\begin{equation}
X\leftrightarrow \theta_\pm ,\qquad \theta_+\leftrightarrow
\theta_- ,
\end{equation}
$F$ as well as $Q_1+Q_2+Q_3$ are invariant. This is the triality
automorphism which may be elegantly formulated in the octonion
formalism \cite{24}. Eq. (13.1) is a particular case of
$Q_1=0=Q_2$, from which, in force of the triality principle, also
$Q_3=P_AP^A=0$ follows.

The relevance of triality for physics and specially for
supersymmetric theories, could be high, however it has not yet
been fully clarified.

For $n=5$, the sixteen component Weyl spinors $\Theta_\pm
=\frac{1}{2} (1+\gg_{11})\Phi$ associated with $\Cl_0(1,9)$ are
simple iff in eq. (2.12) $F_1=0$, that is iff:
\begin{equation}
F_1:=\langle \Theta_\pm^\dagger, G_0G_A\Theta_\pm\rangle =0,\quad
A=1,2,\dots ,10
\end{equation}
where $G_{10} :=\one$.\\
These ten constraint equations have been used for an elegant
super-Poincar\'e covariant description and quantization of
superstrings \cite{25}.

We could also impose simplicity for the 32 component spinors
$\Phi_\pm =\frac{1}{2} (1+\gg_{13})\Omega$ associated with
$\Cl_0(1,11)$, for which in eq.(2.12) $F_2=0$, that is:
\begin{equation}
F_2:=\langle \Phi_\pm^\dagger, \gg_0[\gg_\alpha ,\gg_\beta ]
\Phi_\pm\rangle =0,\quad \alpha, \beta =1,2,\dots ,12
\end{equation}
where $\gg_{12}:=\one$.

These 66 constraint equations might also be relevant for physics since we
know that $\Phi_\pm$ may be conceived as Dirac or Pauli spinors
associated with $\Cl(1.9)$ or with $\Cl(1,10)$ respectively, and if we adopt
our interpretation of the terms in the equations, with indices
$\alpha ,\beta \geq 5$, as interaction-terms, then, among the 66
terms set to zero in eq.(13.6), there are several correlating the
$\Theta_B$-baryon multiplet with the $\Theta_L$-lepton one. In
fact, in the Dirac representation $\gg^{(0)}_\alpha$ of the
generators of $\Cl(1,9)$ $F_2$ implies:
\begin{equation}
\langle \Theta^\dagger_B G_0 G_A G_9 \Theta_L\rangle =0\qquad A= 1,2,\dots ,8
\end{equation}
which if $G_9$ as well as $\G_7$ is assumed in the Dirac
representation, implies that all weak currents between baryons
and leptons are forbidden, and the same may be derived in the
Pauli representation. Then the stability of the lightest baryon:
the proton might have a geometrical explanation in the frame of
simple- or pure-spinor geometry.

\subsection{MINIMAL SURFACES AND STRINGS FROM SPINORS}

\'E. Cartan conjectured \cite{1} that simple spinor geometry may
underlie euclidean geometry insofar null-vectors may be
bilinearly expressed in terms of simple spinors while sums of
null-vectors generally give ordinary euclidean vectors.

In the frame of the main trend of theoretical physics of last
century to identify more and more with geometry and mathematics,
the clue for the implementation of that conjecture might be
envisaged in the central role which null-vectors and null-lines
(lines with null tangent) played in the last two centuries in the
development of geometry in the frame of complex analysis
and which emerged in particular from the Enneper-Weierstrass
\cite{14} parametrization of minimal surfaces in $\R^3$, in the
form
\begin{equation}
X_j(u,v)=X_j(0,0)+{\rm Re} \int^{u+iv}_c Z_j(\alpha ) d\alpha
;\quad j=1,2,3
\end{equation}
where $X_j(u,v)$ are the orthonormal coordinates of the points of
a surface, which is minimal provided $Z_j(\alpha )$ are the
holomorphic coordinates of a null $\C^3$-vector, and $c$ is any
path in the complex plane $(u,v)$ starting from the origin. The
correlation with spinors associated with $\C^3$ are given by
eqs.(3.1), (3.2), where $Z_j=\langle B\varphi
,\s_j\varphi\rangle$ are null. It was shown in ref.[15] that, by
considering $\varphi$ as a Weyl spinor associated with $\Cl(3,1)$
eq.(13.8) may easily be extended to $\R^{3,1}$. For a Majorana
spinor associated with $\Cl(3,1)=R(4)$ the corresponding equation
gives the representation of a string in the form:
\begin{equation}
X_\mu(\s,\tau )=X_\mu(0,0)+ \int^{\s+\tau}_0 t^+_\mu (\alpha )
d\alpha \int^{\s-\tau}_0 t^-_\mu (\beta ) d\beta ;\ \ \ \mu
=0,1,2,3
\end{equation}
where $t^\pm_\mu$ are real, null vectors bilinearly constructed
in terms of Majorana spinors. It was further shown \cite{16} that
the above formalism may be extended to higher dimensional
Clifford algebras and corresponding spinor spaces and that,
whenever Majorana spinors are admitted, strings will be naturally
obtained as integrals of bilinear null vectors in terms of them.
Propositions 1 and 2 of the present paper may be adopted also in
the case of strings such that imbedding simple spinor spaces in
higher dimensional ones implies the corresponding imbedding of
strings. The above may also be extended to momentum space.

In this way the approach presented in this paper could not only
be compatible with, but even be at the origin of string theory;
insofar strings could appear as naturally arising bilinearly,
from (real) spinors through integrations considered as
generalized forms of sums, under certain conditions. It is
interesting to note that the condition of reality and simplicity
for the corresponding spinors brings to the Clifford algebra
\begin{equation}
\Cl(9,1)=\Cl(8,0)+\Cl(1,1)=R(32) ,
\end{equation}
and in this frame gravitation will be naturally contained, as
well known.

A bilinear parametrization of covariant strings and superstrings
theories in terms of Majorana-Weyl spinors associated with
$\Cl(1,9)$ was recently proposed \cite{24}, \cite{25}, \cite{26}.
The general solutions of the equations of motion are obtained
through the use of an octonionic formalism, which might render
transparent, through the triality automorphism, also the
geometrical origin of supersymmetry.

\subsection{FURTHER ASPECTS}

There are certainly several more geometrical aspects which need
to be further analyzed. Among these the role of triality in
supersymmetry theories the possible geometrical origin of
coupling constants in the frame of Riemann geometry, the consequent
formulation of local field theories in the traditional configuration
space and its correlation with momentum space [12] which seems to
be the space where spinor geometry may be naturally formulated
and immediately interpreted as physical equations of motions; a
non surprising fact if one conceives momentum space as the space
of velocities.

Also nullness as well as simplicity constraints should be further
analyzed since, while introducing the elegant conception of
projective manifolds and of compact momentum spaces, they could
also furnish sum rules of physical meaning.

\section{CONCLUSIONS}

Fermion multiplets may be represented by spinor multiplets and,
from these, if simple or pure, following \'E. Cartan, vectors of
null quadrics in pseudo euclidean spaces may be bilinearly
constructed. It is then natural to tray to imbed spinor spaces,
and the corresponding bilinearly constructed vector spaces, in
higher dimensional ones starting from the most elementary non
trivial one: the two-component Pauli spinors of ordinary
3-dimensional space associated with $\Cl(3)$. It is remarkable
that in this construction the signature of the constructed
spaces, and of their Clifford algebras, results unambiguously
defined starting from $\Cl(3,1)$ or $\Cl(1,3)$ up to $\Cl(9,1)$
or $\Cl(1,9)$, after which the sequence is repeated.

It might be not accidental that, on the way of this construction,
most of the elementary equations of motion of fermion-physics
(including Weyl, and from these, also Maxwell's equations) are
naturally found, however in momentum space or the space of
velocities, which naturally results compact; and that these equations naturally
present some of
the main groups of internal symmetry like $U(1),SU(2)$ and
$SU(3)$, each one deriving from the 3 degree of complexity of the
field of numbers; that is: complex numbers, quaternions and
octonions, respectively. This result may be correlated with the general
properties of the division algebra like the real $\Cl(1,9)$ to which
we arrived in our construction as shown by G. Dixon [24].

In this preliminary approach we concentrated our attention on
some of the aspects of the geometrical construction  and on the possible
role of the elegant concept of simple or pure spinors, but much
more has to be done specially in the study of its possible
relevance for physics, and of the correlation of the resulting
compact momentum spaces with figuration space, which seems to be
necessary only up to dimension four, the realm of Weyl, Maxwell
and Dirac equations and of local field. Beyond four, dimensional
reduction appears natural in momentum spaces, where it simply
eliminates interaction terms from the equations of motion, while
it appears redundant in the corresponding configuration space,
for which there seems to be no need, for dimensions larger than
four.

\end{document}